\documentclass[a4paper,12pt]{article}
\usepackage[a4paper, margin=1in]{geometry}
\usepackage[american]{babel}
\usepackage{latexsym}
\usepackage{float}
\usepackage{amsmath}
\usepackage{amssymb}
\usepackage{caption}
\usepackage[dvips]{graphicx}
\usepackage[dvipsnames]{xcolor}
\usepackage{bm}
\usepackage{booktabs}
\usepackage{fancyvrb}
\usepackage{natbib}
\usepackage{setspace}

\newcommand{\norm}[1]{\left\lVert#1\right\rVert}
\DeclareMathOperator*{\argmin}{arg\,min}
\newenvironment{keywords}{\noindent\textbf{keywords:}}{}
\title{\textbf{Adaptive sparse group LASSO in quantile regression}}
\author{Álvaro Méndez Civieta\thanks{Department of Statistics, University Carlos III of Madrid.}\,{\ }\thanks{uc3m-Santander Big Data Institute.} \and M. Carmen Aguilera-Morillo\thanks{Department of Applied Statistics and Operational Research, and Quality, Universitat Politècnica de València}\,{\ }\footnotemark[2] \and Rosa E. Lillo\footnotemark[1]\,{\ }\footnotemark[2]}
\date{}
\begin{document}
\maketitle
\vspace{-0.5in}
\begin{abstract}
	 This paper studies the introduction of sparse group LASSO (SGL) to the quantile regression framework. Additionally, a more flexible version, an adaptive SGL is proposed based on the adaptive idea, this is, the usage of adaptive weights in the penalization. Adaptive estimators are usually focused on the study of the oracle property under asymptotic and double asymptotic frameworks. A key step on the demonstration of this property  is to consider adaptive weights based on a initial $\sqrt{n}$-consistent estimator. In practice this implies the usage of a non penalized estimator that limits the adaptive solutions to low dimensional scenarios. In this work, several solutions, based on dimension reduction techniques PCA and PLS, are studied for the calculation of these weights in high dimensional frameworks. The benefits of this proposal are studied both in synthetic and real datasets.
\end{abstract}


\section{Introduction}\label{sec:intro}
Along years, regression has become a key method in statistics. Least squares (LS) regression estimates the conditional mean response of a variable as a function of the covariates.  Usually, these models assume the errors to be centered, homoscedastic and independent. Making this assumptions, it is guaranteed that the LS estimator is the best linear unbiased estimator, or a BLUE estimator. Additionally, if the errors are assumed to be Gaussian one can perform finite sample studies. However, these hypothesis are not always verified in practical applications, and the LS estimator is known to be extremely sensitive to  the presence of outliers or heavy tailed distributions, making it perform poorly when the errors are non Gaussian. Ever since the seminal work of \cite{Koenker1978}, quantile regression (QR) models have gained importance when dealing with this kind of situations. QR models allow for a relaxation of the classical first two moment conditions over the model error.  In addition, the errors in QR are not required to be Gausian. This means that QR offers robust estimators capable of dealing with heteroscedasticity and outliers. QR models can also estimate different quantile levels of a response variable, giving a precise insight of the relation between response and covariates at upper and lower tails. This can provide a much richer point of view than OLS regression. For a full review on quantile regression, we recommend \citep{Koenker2005}.

In recent years, high dimensional data in which the number of covariates $p$ is larger than the number of observations $n$ $(p\gg n)$, has become increasingly common. This problem can be found in many different areas like computer vision and pattern recognition \citep{Wright2010}, climate data over different land regions \citep{Chatterjee2011}, and prediction of cancer recurrence based on patients genetic information \citep{Simon2013}, \citep{YahyaAlgamal2019}. In these scenarios, variable selection gains in special importance offering sparse modeling alternatives that help identifying significant covariates and enhancing prediction accuracy. One of the first and most popular sparse regularization alternatives is LASSO, which was proposed by \cite{Tibshirani1996} and adapted to the QR framework by \cite{Li2008}, who developed the piece-wise linear solution of this technique. LASSO is a technique that penalizes each variable individually, enhancing thus individual sparsity. However, in many real applications variables are structured into groups, and group sparsity rather than individual sparsity is desired. One can think for example of a genetic dataset grouped into gene pathways. This problem was faced by the group LASSO penalization of \cite{Yuan2006}, and opened the doors to more complex penalizations like the sparse group LASSO \citep{Friedman2010}, which is a linear combination of LASSO and group LASSO providing solutions that are both between and within group sparse. With the same objective in mind, \cite{Zhou2010} proposed a hierarchical LASSO.  Other studies have worked on properties for robust estimators in regression when the number of covariates increase with sample size (see for example \cite{Huber2009}). In the same line, it is also worth mentioning the work from \cite{Loh2017}, that extends the usage of robust estimators, like those obtained using Hubert or Tuckey loss functions (among others) to high dimensional settings, introducing a set of generalized M-estimators capable of dealing with outliers in both the errors and the covariates terms. To the best of our knowledge, the SGL technique has not been studied in the framework of QR models, so this gap is addressed first, extending the SGL penalization to quantile regression.

\cite{Zou2006a} was the first to propose the usage of  adaptive weights for each variable on the LASSO penalization as a way to increase the model flexibility and correct the estimator bias. This idea, generally known as the adaptive idea, was then extended to other penalizations. The weights of the adaptive idea are defined in the literature based on an initial  $\sqrt{n}$-consistent estimator. Typically, this is the result of a nonpenalized model. This definition is a key step for the demonstration of  the oracle property of the estimators (in the sense of \cite{Fan2001}), but it is also restrictive, as it limits the usage of adaptive penalizations just to the  situations in which solving a nonpenalized model is a feasible first step. This approach, focused on the oracle property under asymptotic, or even double asymptotic frameworks is observed in \cite{Nardi2008} for the adaptive group LASSO, \cite{Ghosh2011} for an adaptive elastic net, \cite{Ciuperca2019} for the adaptive group LASSO in QR, \cite{Ciuperca2017} for the adaptive fused LASSO  in QR, \cite{Wu2009} for the adaptive LASSO and SCAD penalizations in QR, and \cite{Zhao2014} for an adaptive hierarchical LASSO in QR among others. It is especially interesting to remark the work developed by \cite{Poignard2018}, in which an adaptive sparse group LASSO estimator suitable for low dimensional scenarios (with $n>p$) is proposed, studying its theoretical properties for a set of general convex loss functions. 

The main contribution of this work lies here. An adaptive sparse group LASSO (ASGL) for quantile regression estimator is defined, working especially on enabling the usage of the ASGL estimator in high dimensional scenarios (with $p\gg n$). In order to achieve this objective, four alternatives for the weight calculation step are proposed. It is worth noting that these weight calculation alternatives can be used not only in the case of the ASGL estimator, but also in the rest of the adaptive-based estimators available in the literature. The performance of these alternatives is also studied in the case of low dimensional scenarios, making the proposed work a good alternative for both high dimensional and low dimensional problems.

The rest of the paper is organized as follows. In Section \ref{sec:pen} some basic theoretical concepts are introduced, along with the formal definition of the sparse group LASSO in quantile regression. This definition is extended to the adaptive idea in Section \ref{sec:asgl_qr}, proposing the ASGL estimator. Section \ref{sec:oracle_prop} discusses the main results regarding asymptotic behavior of adaptive estimators, and Section \ref{sec:adaptive_weights} introduces the weights calculation alternatives for high dimensional scenarios, as well as some remarks regarding the asymptotic behavior of the proposed alternatives. Simulation results are divided into two blocks: Section \ref{sec:sym_simetric_error} shows the advantages of this proposal in synthetic datasets in high and low dimensional scenarios considering a symmetric error distribution while the supplementary material shows a sensitivity analysis of the proposed methods under skewed distribution errors as well as the effect of different hyperparameter values. In Section \ref{sec:real_data} the proposed model is used in a real dataset, a genomic dataset including gene expression data of rat eye disease first shown in \cite{Scheetz2006}. The computational aspects of the problem are briefly commented in Section \ref{sec:comp}, and the conclusions are provided in Section \ref{sec:conclusion}. 


\section{Penalized quantile regression}\label{sec:pen}
Consider a sample of $n$ observations structured as $\mathbb{D}=(y_i, \bm{x}_i),{\ }i=1,\ldots,n$ from some unknown population and define the following linear model,
\begin{equation}\label{eq:lm}
y_i = \bm{x}_i^t\bm{\beta}+\varepsilon_i,{\ }i=1,\ldots,n
\end{equation}
where $y_i$ is the i-\emph{th} observation of the response variable, $\bm{x}_i\equiv(x_{i1},\ldots,x_{ip})$ is the vector of $p$ covariates for observation $i$ and $\varepsilon_i$ is the error term.

Let us introduce now the quantile regression framework by defining the loss check function,
\begin{equation}\label{eq:check_function}
\rho_{\tau}(u)=u(\tau-I(u<0))
\end{equation} 
where $I(\cdot)$ is the indicator function. In their seminal work \cite{Koenker1978} proved that the $\tau$-\emph{th} quantile of the response variable can be estimated by solving the following optimization problem,
\begin{equation}\label{eq:qr}
\tilde{\bm{\beta}}=\argmin_{\bm{\beta}\in\mathbb{R}^p}\left\{R(\bm{\beta})\right\}.
\end{equation}
where $R(\bm{\beta})$ defines the risk function of quantile regression,
\begin{equation}\label{eq:risk_qr}
R(\bm{\beta})=\frac{1}{n}\sum_{i=1}^n\rho_{\tau}(y_i-\bm{x}_i^t\bm{\beta})
\end{equation}

Quantile regression models allow for a relaxation of the classical first two moment conditions over the model errors $\varepsilon_i$ defined in equation \ref{eq:lm}. These errors are no longer required to be centered, homoscedastic or normally distributed, as stated in \cite{Koenker2005}, offering robust estimators capable of dealing with heteroscedasticity and outliers. 

We call high dimensional scenarios to the datasets in which $p$ is much larger than $n$ ($p\gg n$). This problem is becoming more and more common nowadays, and can be observed in many different fields of research such as computer vision and pattern recognition \citep{Wright2010}, climate data over different land regions \citep{Chatterjee2011} or prediction of cancer recurrence based on patients genetic information \citep{Simon2013}. An alternative that has been intensively studied in recent years for dealing with these scenarios is the penalization approach. By penalizing a regression model it is possible to perform variable selection and improve the accuracy and interpretability of the models.

One of the best known variable selection penalization methods is the least absolute selection and shrinkage operator, generally known as LASSO, proposed initially by \cite{Tibshirani1996} which, in the case of the QR framework solves,
\begin{equation}\label{eq:qr_l}
\hat{\bm{\beta}}=\argmin_{\bm{\beta}\in\mathbb{R}^p}\left\{R(\bm{\beta})+\lambda\norm{\bm{\beta}}_1\right\},
\end{equation}
where $R(\bm{\beta})$ is the QR risk function defined in equation \eqref{eq:risk_qr}. The LASSO penalization sends many $\bm{\beta}$ components to zero, offering sparse solutions and performing automatic variable selection. In the last years, many LASSO-based algorithms have been proposed. \cite{Yuan2006} introduced the group LASSO penalization as an answer for the need to select variables not individually but at the group level. This penalization solves the following problem,
\begin{equation}\label{eq:qr_gl}
\hat{\bm{\beta}}=\argmin_{\bm{\beta}\in\mathbb{R}^p}\left\{R(\bm{\beta})+\lambda\sum_{l=1}^K\sqrt{p_l}\norm{\bm{\beta}^l}_2\right\},
\end{equation}
where $K$ is the number of groups, $\bm{\beta}^l\in\mathbb{R}^{p_l}$ are vectors of components of $\bm{\beta}$ from the l-\emph{th} group, and $p_l$ is the size of the l-\emph{th} group. The group LASSO penalization works in a similar way to LASSO, but while LASSO enhances sparsity at individual level, group LASSO enhances sparsity at group level, selecting, or sending to zero whole groups of variables.

Initially proposed by \cite{Friedman2010}, the sparse group LASSO (SGL) is a linear combination of LASSO and group LASSO penalizations. Well known in linear regression and other GLM models, to the best of our knowledge SGL has not been adapted to QR, and as a first step in the paper, this penalization is introduced.
\begin{equation}\label{eq:qr_sgl}
\hat{\bm{\beta}}=\argmin_{\bm{\beta}\in\mathbb{R}^p}\left\{R(\bm{\beta})+\alpha\lambda\norm{\bm{\beta}}_1+(1-\alpha)\lambda \sum_{l=1}^K\sqrt{p_l}\norm{\bm{\beta}^l}_2\right\}.
\end{equation}
As in LASSO and group LASSO, SGL solutions are, in general, sparse, sending many of the predictor coefficients to zero. However, while LASSO solutions are sparse at individual level, and group LASSO solutions are sparse at group level, SGL offers both between and within group sparsity, outperforming both alternatives.

\begin{figure}[]
	\centering
	\caption{Contour lines for LASSO, group-LASSO and sparse-group-LASSO penalties in the case of a single 2-dimensional group}
	\includegraphics[width=9cm]{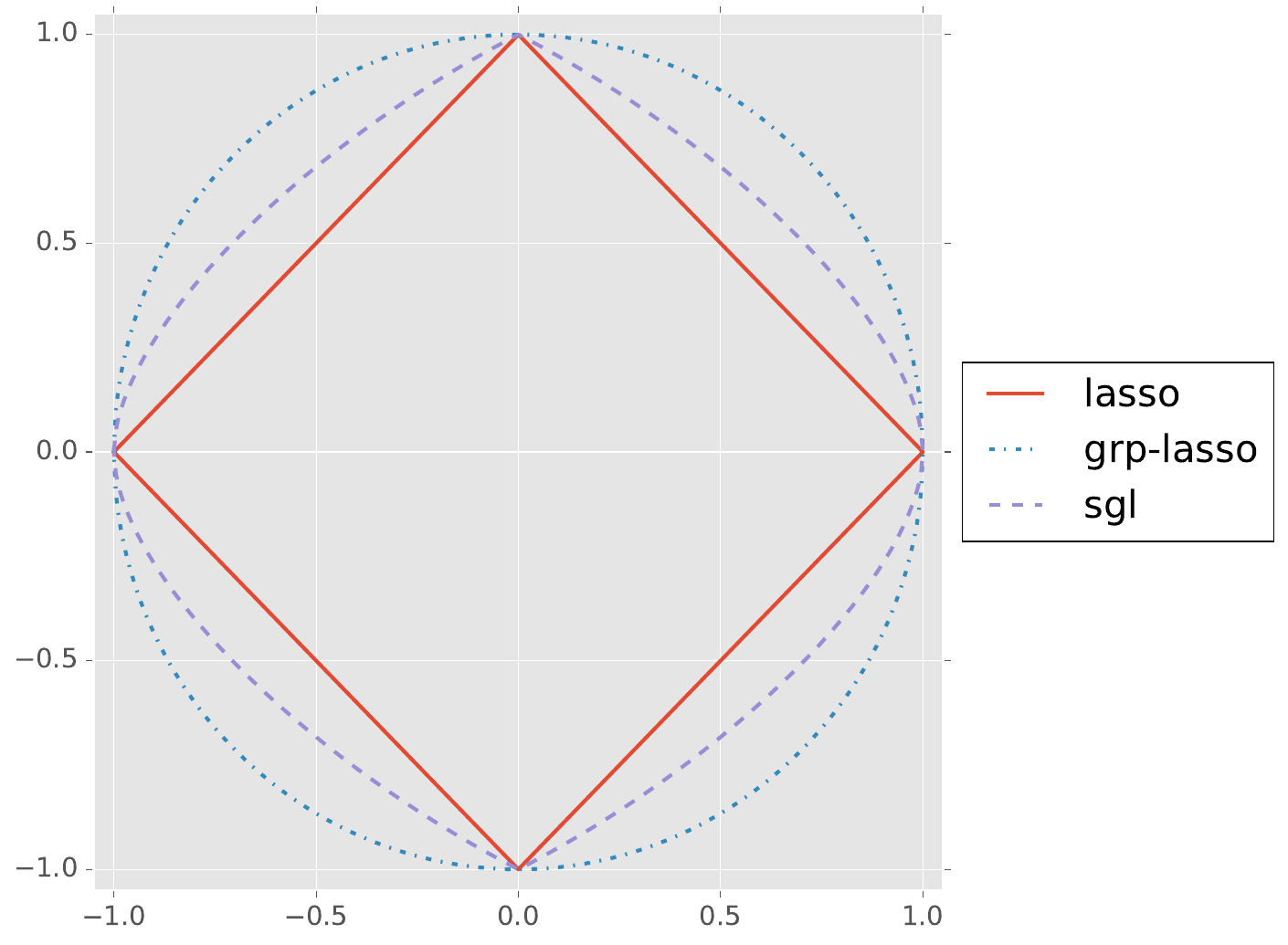}
	\label{im:sgl_comparison}
\end{figure}

From an optimization perspective, equation \eqref{eq:qr_sgl} defines a sum of convex functions. This convexity ensures that the solution of the minimization problem is a global minimum. Figure \ref{im:sgl_comparison} shows the constrains defined by LASSO, group LASSO and SGL in the case of a single 2-dimensional group of predictors. 


\section{Adaptive sparse group LASSO}\label{sec:asgl_qr}
From an empirical perspective, sparse group LASSO shows great performance. However, due to its mathematical formulation,  it applies a constant penalization rate that  provides  biased estimates for large coefficients. The adaptive idea, initially introduced by \cite{Zou2006a} is considered here as a way to correct this limitation. In this work, a variant of the SGL penalization, the adaptive sparse group LASSO (ASGL) for quantile regression is defined. The ASGL estimator for QR is the result of the following minimization process,
\begin{equation}\label{eq:asgl_qr}
\hat{\bm{\beta}}=\argmin_{\bm{\beta}\in\mathbb{R}^p}\left\{R(\bm{\beta})+\alpha\lambda\sum_{j=1}^p\tilde{w}_j\lvert\beta_j\rvert+(1-\alpha)\lambda\sum_{l=1}^K\sqrt{p_l}\tilde{v_l}\norm{\bm{\beta}^l}_2\right\},
\end{equation}
where $\tilde{\bm{w}}\in\mathbb{R}^p$ and $\tilde{\bm{v}}\in\mathbb{R}^K$ are known weights vectors and $R(\bm{\beta})$ is the risk function for quantile regression defined in equation \ref{eq:risk_qr}. The intuition behind these weights is that if a variable (or group of variables) is important, it should have a small weight, and this way would be lightly penalized. On the other hand, if it is not important, by setting a large weight it is heavily penalized. This enhances the model flexibility and improves variable selection and prediction accuracy. It is worth saying that this formulation defines a convex function and thus, the global minimum can be found.


	\section{The oracle property}\label{sec:oracle_prop}
	An estimator is oracle if it can correctly select the nonzero coefficients in a model with probability converging to one, and if the nonzero coefficients are asymptotically normally distributed. These properties were initially defined in \cite{Fan2001}, where they proved that the SCAD was an oracle estimator under an asymptotic framework of fixed dimension $p$. The oracle property of the SCAD estimator was then extended in \cite{Fan2004} to a double asymptotic framework of $p$ depending on $n$. This is, $p\rightarrow \infty$ as $n\rightarrow \infty$, but $p$ growing at a lower rate and always $n>p$. \cite{Zou2006a} proved that the LASSO was not an oracle estimator due to the bias generated by the constant penalization rate. They proposed the usage of adaptive weights as a means to correct the bias, showing that the adaptive LASSO was an oracle estimator under the asymptotic framework of fixed $p$, as long as the weights required by the adaptive idea were computed based on a initial $\sqrt{n}$-consistent estimator. Actually, they proposed using the result from a non penalized model for the computation of the weights $\tilde{\bm{w}}$,
		\begin{equation}
		\tilde{w_i}=\frac{1}{\lvert\tilde{\beta_i}\rvert^{\gamma}},
		\end{equation}
		where $w_i$  and $\tilde{\beta_i}$ correspond to the  i-\emph{th} element of vectors $\tilde{\bm{w}}$ and $\tilde{\bm{\beta}}$ respectively, $\lvert\cdot\rvert$ denotes the absolute value function, $\gamma$ is a non negative constant and $\tilde{\bm{\beta}}$ is the solution vector obtained from the unpenalized model (described, in the case of the QR framework, in equation \eqref{eq:qr}). 

	Ever since then, the adaptive idea has been extended to many LASSO-based formulations in OLS, GLM and QR models among others. One can see for instance \citep{Ghosh2011} where an adaptive elastic net is defined, \citep{Wu2009} that introduces the adaptive LASSO in QR, \citep{Ciuperca2017} where an adaptive fused LASSO in QR is defined, \citep{Zhao2014} who proposes an adaptive hierarchical LASSO in QR or \citep{Poignard2018}, where an adaptive sparse group LASSO estimator is defined in a general set of convex functions, among others. All these works are centered on the demonstration of the oracle property under the asymptotic or double asymptotic framework, being the usage of an initial $\sqrt{n}$-consistent estimator on the calculations of the weights a key step in the demonstration. A major drawback of this approach in our opinion is precisely that the asymptotic or double asymptotic frameworks are limited to low dimensional scenarios where $n>p$ but do not consider high dimensional scenarios where $p\gg n$. This is remarked by the fact that usually, the initial $\sqrt{n}$-consistent estimators used in the weight calculations are taken from non penalized models, only feasible in low dimensional scenarios.
	
	 Dealing with the problem of an increasing number of covariates is, however, challenging. When an OLS model is considered, the third order term of the taylor expansion on the loss function vanishes, but out of this framework, for example in GLM or QR models, this term does not vanish, and additional boundaries on the convergence rates of $p$ (the number of variables) and $n$ (the number of observations) are required in order to demonstrate the consistency and the oracle property of the estimators. This is pointed out in detail, for a general framework of convex functions, in \cite{Poignard2018}.
	
	When considering a high dimensional scenario it is possible to find very interesting results from recent years. One can see for example \citep{Huang2008}, who considers the oracle property of a bridge penalized least squares model under the $p\gg n$ framework as long as the bridge parameter is strictly between $0$ and $1$ (leaving out of the formulation the LASSO estimator). In order to achieve these results, they require additional conditions on the design matrix $X$, namely, they require partial orthogonality between the set of significant variables and the set of non significant variables. Similar results can be observed for the adaptive LASSO in least squares \citep{Huang2008b} where partial orthogonality conditions are required to demonstrate the oracle property in high dimensions, for the SCAD penalization in linear models in \cite{Kim2008} and for the SCAD and MCP penalizations in quantile regression in \cite{Wang2012}.  However, the conditions required on the design matrix (and therefore on the covariates) to fit the oracle property are difficult to verify in practice. Thus, the results have an important mathematical relevance that should be landed in more realistic hypotheses.


\section{Adaptive weights calculation}\label{sec:adaptive_weights}
The objective of this section is to introduce different alternatives for the calculation of weights in the adaptive framework. The intuitive idea is to find a way to substitute $\tilde{\beta}$, the solution from the unpenalized model, unfeasible in high dimensional scenarios, in the calculation of the adaptive weights. This problem will be faced making use of two dimensionality reduction techniques, principal component analysis (PCA) and partial least squares (PLS). The proposed weight calculation alternatives can be used both in high dimensional and low dimensional scenarios. It is worth highlighting that these alternatives can be applied not only to the ASGL algorithm, but also to other adaptive based algorithms.
\subsection{Principal components analysis}\label{sec:weigths_pca}
Given the covariates matrix $\bm{X}\in\mathbb{R}^{n\times p}$ defined in equation \eqref{eq:lm}, with maximum rank $r=\min{\left\{n,p\right\}}$, consider the matrix of principal components $\bm{Q}\in\mathbb{R}^{p\times r}$ defined in a way such that the first principal component has the largest possible variance, and each succeeding component has the largest possible variance under the constraint that it is orthogonal to the preceding components. From an algebra perspective, the principal components in $\bm{Q}$ define an orthogonal change of basis matrix that maximize the variance explained from $\bm{X}$. Consider $\bm{Z}=\bm{XQ}\in\mathbb{R}^{n\times r}$ the projection of $\bm{X}$ into the principal components subspace. Two weight calculation alternatives based on principal components are proposed.
\subsubsection{Based on a subset of components}\label{sec:weight_pca_subset}
Consider the submatrix $\bm{Q}_d=[\bm{q}_1,\ldots,\bm{q}_d]^t$ where $\bm{q}_i\in\mathbb{R}^p$ is the i-\emph{th} column of the matrix $\bm{Q}$, and $d\in\left\{1,\ldots,r\right\}$ is the number of components chosen. Let $\alpha_{pca,d}\in[0,100]$ be the percentage of variability from $\bm{X}$ that the principal components in $\bm{Q}_d$ are able to explain. If $d=r$ then the principal components in $\bm{Q}_d$ are able to explain all the original variability from $\bm{X}$, and $\alpha_{pca,d}=100$. If $d<r$ then $\alpha_{pca,d}<100$. The number of components chosen in order to explain up to a certain percentage of variability is fixed by the researcher. Obtain $\bm{Z}_d=\bm{XQ}_d\in\mathbb{R}^{n\times d}$ the projection of $\bm{X}$ into the subspace generated by $\bm{Q}_d$ and solve the unpenalized model,
\begin{equation}\label{eq:pca_pct}
\tilde{\bm{\beta}}=\argmin_{\bm{\beta}\in\mathbb{R}^d}\left\{\frac{1}{n}\sum_{i=1}^n\rho_{\tau}(y_i-\bm{z}_i^t\bm{\beta})\right\}.
\end{equation}
This model defines a low dimensional scenario where $\tilde{\bm{\beta}}\in\mathbb{R}^d$. Using this solution, it is possible to obtain an estimation of the high dimensional scenario solution, $\hat{\bm{\beta}}=\bm{Q}_d\tilde{\bm{\beta}}\in\mathbb{R}^p$. Finally, the weights are estimated as,
\begin{equation}\label{eq:weight_pca_subset}
\tilde{w_j}=\frac{1}{\lvert\hat{\beta}_j\rvert^{\gamma_1}} {\ } \textup{ and } {\ } \tilde{v_l}=\frac{1}{\norm{\hat{\bm{\beta}}^l}_2^{\gamma_2}},
\end{equation}
where $\hat{\beta}_j$ is the j-\emph{th} component from $\hat{\bm{\beta}}$, $\hat{\bm{\beta}}^l$ is the vector of components of $\bm{\beta}$ from the l-\emph{th} group, and $\gamma_1$ and $\gamma_2$ are non negative constants usually taken in $[0,2]$.
\subsubsection{Based on the first component}
A more straightforward approach based on the first principal component is also proposed. The principal components are no more than linear combinations of the original variables. Therefore, the first principal component $\bm{q}_1\in\mathbb{R}^p$, which is the first column of the matrix $\bm{Q}$, includes one weight for each of the $p$ original variables. This proposal consists of calculating the weights as,
\begin{equation}\label{eq:weight_pca_first_comp}
\tilde{w_j}=\frac{1}{\lvert q_{1j}\rvert^{\gamma_1}} {\ } \textup{ and } {\ } \tilde{v_l}=\frac{1}{\norm{\bm{q}_1^l}_2^{\gamma_2}},
\end{equation}
where $q_{1j}$ is the j-\emph{th} component from $\bm{q}_1$ and defines the weight associated to the j-\emph{th} original variable, $\bm{q}_1^l$ is the vector of components of $\bm{q}_1$ from the l-\emph{th} group and $\gamma_1$ and $\gamma_2$ are non negative constants usually taken in $[0,2]$.
\subsection{Partial least squares}\label{sec:weight_pls_subset}
The principal components are defined in a way such that they capture the maximum possible variance from $\bm{X}$ under the constraint that they are orthogonal to the rest of the principal components. However, being relevant for describing the variance of $\bm{X}$ does not necessarily mean that a principal component is relevant for predicting the value of $\bm{y}$. Partial least squares (PLS) is a dimensionality reduction technique centered on maximizing the covariance between $\bm{X}$ and $\bm{y}$.

Given the covariates matrix $\bm{X}\in\mathbb{R}^{n\times p}$ defined in equation \eqref{eq:lm}, with maximum rank $r=\min{\left\{n,p\right\}}$, consider the matrix of PLS components $\bm{T}\in\mathbb{R}^{p\times s}$ and the projection of $\bm{X}$ into the subspace generated by $\bm{T}$: $\bm{U}=\bm{XT}\in\mathbb{R}^{n\times s}$. The matrix of PLS components $\bm{T}$ defines a nonorthogonal change of basis matrix whose projection $\bm{U}$ is computed in a way such that the first projection vector, $\bm{u_1}\in\mathbb{R}^n$ has the largest possible covariance with $\bm{y}$, and each succeeding projection vector has the largest possible covariance with $\bm{y}$ under the constraint that it is uncorrelated to the rest of the projection vectors.

Given the submatrix $\bm{T}_d=[\bm{t}_1,\ldots,\bm{t}_d]^t$ where $\bm{t}_i\in\mathbb{R}^p$ is the i-\emph{th} column of the matrix $\bm{T}$, and $d\in\left\{1,\ldots,s\right\}$ is the number of components chosen, let $\alpha_{pls,d}\in[0,100]$ be the percentage of variability from $\bm{X}$ that the PLS components in $\bm{T}_d$ are able to explain. The nonorthogonality of $\bm{T}$ implies that the total number of PLS components available to be computed is smaller than the rank of $\bm{X}$, $s\leq r$, and that the maximum possible percentage of variability explained by the PLS components $\alpha_{pls,s}$ is then lower than $100\%$.

In the case of principal components analysis, the matrix of principal components $\bm{Q}$ defines an orthogonal change of basis matrix that results into an orthogonal projection matrix $\bm{Z}$ maximizing the variance of $\bm{X}$. On the other hand, PLS defines a nonnecesarily orthogonal change of basis matrix $\bm{T}$ that results into an uncorrelated projection matrix $\bm{U}$ maximizing the covariance between $\bm{U}$ and $\bm{y}$. In the same way as for the PCA alternatives proposed, two alternatives of weight calculation using PLS are  considered: based on a subset of PLS components, and based just on the first PLS component.


	\subsection{Influence of PCA and PLS on the oracle property}\label{sec:oracle_proerties_wc}
	As commented in Section \ref{sec:oracle_prop}, a key condition in the demonstration of the oracle property in adaptive estimators is to assume that the initial estimator used in the weights calculation is $\sqrt{n}$-consistent. 
	
	The usage of $pca_d$ or $pls_d$ weight calculation proposes to consider a subset of $d$ components in the estimation of the weights. A question that may arise here is whether these PCA (or PLS) estimator is $\sqrt{n}$-consistent or not. We propose the following simple low dimensional example in the OLS framework that can help answering this question.\\{\ }\\\noindent
		\textit{Example:}\\
	Given the random variables $X_1\sim N(0, 0.99)$ and $X_2\sim N(0, 0.01)$, consider the random vector : $X=(X_1, X_2)$, for which
	\begin{equation*}
	cov(X)= 
	\begin{pmatrix}
	0.99 & 0\\
	0 & 0.01
	\end{pmatrix}.
	\end{equation*}
	And thus, the eigenvalues from cov(X) are $\lambda_1=0.99$ and $\lambda_2=0.01$, and the matrix of eigenvectors  is
	\begin{equation*}
	P= 
	\begin{pmatrix}
	1 & 0\\
	0 & 1
	\end{pmatrix}.
	\end{equation*}
	If PCA is applied on this random vector $X$, the rotation matrix obtained will be $P$, yielding to a first principal component that explains $99\%$ of the original variability and a second principal component that explains the remaining $1\%$.
	
	Consider now the following linear model,
	\begin{equation*}
	y=X\beta+\varepsilon,
	\end{equation*} 
	where $\beta=(0,100)^t$ and $\varepsilon\sim N(0,0)$. Following the steps described in Section \ref{sec:weight_pca_subset}, consider a subset of components that explain up to a certain percentage of variability, for example, $99\%$ of the variability. This implies that $X$ will be projected onto the subspace spanned just by the first principal component $P_1$, $Z=XP_1=X_1$. Solve now the linear model $\tilde{y}=Z\tilde{\beta},$ where
	$$\tilde{\beta}=\dfrac{cov(Z,y)}{var(Z)}=\dfrac{cov(X_1,y)}{var(X_1)}=0.$$
	Then, the projection of the estimator $\tilde{\beta}$ into the original subspace is given by $\hat{\beta}=P_1\tilde{\beta}=(0,0)^t$. Now, in order to be $\sqrt{n}$-consistent, an estimator  should verify:
	\begin{equation*}
	(\hat{\beta}-\beta) \;\textup{is}\; O_p(n^{-1/2}) \;\textup{if}\; \textup{for all} \;\varepsilon>0 \; \exists K>0 \;\textup{s.t.}\;
	\end{equation*}
	\begin{equation*}
	P_{n\rightarrow\infty}(\sqrt{n}|\hat{\beta}-\beta|>K)<\varepsilon
	\end{equation*}
	Taking into account that $\beta=(0,100)^t$, it is clear that the $\sqrt{n}$-consistency property is not verified by $\hat\beta$. The problem arises because the variability in variable $Y$ is explained by $X_2$, which is not selected because it explains only $1\%$ of the total variability of $X$.
		
		We would like to point out that this example is meant to be a counterexample of a situation in which the $pca_d$ is not $\sqrt{n}$-consistent. However, in our opinion, it clarifies the conditions required by the estimator in order to be consistent, as stated in the following remarks.
	\\
	
	\noindent\textit{Remark 1}. Consider an ASGL estimator, where the weights are computed based on a subset of principal components $pca_d$ in the asymptotic or double asymptotic frameworks. If all the components are selected (this is, if the components explain $100\%$ of the original variability), then the initial estimator used in the weights calculation is $\sqrt{n}$-consistent, and therefore, the ASGL estimator is an oracle estimator. Observe that by selecting all the components, $\hat\beta=Q\tilde\beta$ is equal to the unpenalized estimator defined in equation \eqref{eq:qr}.
	\\
	
	\noindent\textit{Remark 2}. As shown in Section \ref{sec:oracle_prop}, the proof of the oracle property of an estimator in high dimensional scenarios is much more complex than in low dimensional scenarios. We conjecture that in the high dimensional context, the $pca_d$ estimator will behave in a similar way as in low dimensional scenarios, requiring to achieve a $100\%$ of explained variability, but requiring also additional hypothesis similar to the ones observed in, for example, \cite{Wang2012}. In this paper, a set of $5$ previous conditions is required for the demonstration of the oracle property in a high dimensional framework in quantile regression while considering non convex penalizations (such as SCAD). Among other things, the proposed conditions include restrictions on the design matrix, for example, that given the design matrix $\bm{X}$, $\bm{S}=\frac{1}{n}\bm{X^t}\bm{X}$ should be bounded, and the eigenvalues of $\bm{S}$ should be bounded as well. We consider that due to the complexity of the required results, studying the theoretical aspect of the estimator in high dimensional scenarios is a topic for further work. However, we study the behavior of this estimator in high dimensional scenarios both in synthetic and real datasets in Sections \ref{sec:sym_simetric_error} and \ref{sec:real_data}, and in the supplementary material, obtaining very good results.
	\\
		
	\noindent\textit{Remark 3}. The study of the oracle property of the $pls_d$ estimator is much more complex than this of $pca_d$. As commented in section \ref{sec:weight_pls_subset}, the maximum percentage of variability explained by the PLS components can be smaller than $100\%$, and thus, we would be facing the same issues described in the example above. This situation will also be a topic for further work.

	\section{Simulation study: symmetric errors}\label{sec:sym_simetric_error}
	This section shows the performance of the proposed ASGL estimator under different synthetic dataset examples focused on symmetric errors as it is usual in OLS models. The proposed ASGL estimator is studied here under the framework of the following model,
	\begin{equation*}
	y=X\beta+\varepsilon,  {\ }\varepsilon\sim t(3),
	\end{equation*}

where the data matrix $X$ is generated from a standard Gaussian distribution. Variables are organized in groups, considering a within group correlation of $0.5$ and a between group correlation of $0$. A quantile level $\tau=0.5$ is considered. The scheme used here is an adaptation of other simulation schemes used in \cite{Wu2009} and \cite{Zhao2014}.

Given that the ASGL formulation in equation \eqref{eq:asgl_qr} includes a weight penalization on the group LASSO part based on the group size (the term $\sqrt{p_l}$), two model formulations are considered: 
\begin{itemize}
	\item Adaptive LASSO in sparse group LASSO (AL-SGL), where $\tilde{\bm{w}}\neq\bm{1}$ but $\tilde{\bm{v}}=\bm{1}$, in which the adaptive idea is only applied to the LASSO part.
	\item Adaptive sparse group LASSO (ASGL), where $\tilde{\bm{w}}\neq\bm{1}$ and $\tilde{\bm{v}}\neq\bm{1}$.
\end{itemize} 
Furthermore, the four weight calculation alternatives proposed are studied:
\begin{itemize}
	\item PCA weights based on regression on a subset of principal components, we denote this as $pca_{d}$;
	\item PCA weights based on the first principal component, we denote this as $pca_{1}$;
	\item PLS weights based on regression on a subset of PLS components, we denote this as $pls_{d}$;
	\item PLS weights based on the first PLS component, we denote this as $pls_{1}$.
\end{itemize}

The total number of components $d$ used in the weight estimation in $pls_d$ and $pca_d$ is chosen such that in both cases the percentage of variability explained from the original matrix $\bm{X}$ is $\alpha_{pca,d}=80\%$, $\alpha_{pls,d}=80\%$. As commented along Section \ref{sec:adaptive_weights}, due to the non orthogonality of the PLS components it can happen that the maximum possible variability explained by the PLS components $\alpha_{pls,s}$ is smaller than $80\%$. In these cases we consider $d$ such that $\alpha_{pls,d}=\alpha_{pls,s}$.

The results obtained by the models proposed in this work are compared with the results from LASSO and SGL formulations. For each dataset $\mathbb{D}$, a partition into three disjoint subsets, $\mathbb{D}_{train}$, $\mathbb{D}_{val}$ and $\mathbb{D}_{test}$ is considered. $\mathbb{D}_{train}$ is used for training the models, this is, solving the model equations. $\mathbb{D}_{val}$ is used for validation, this is, optimizing the model parameters. This optimization is performed based on grid-search. Finally, $\mathbb{D}_{test}$ is used for testing the models prediction accuracy. The model parameters are optimized based on the minimization of the quantile error, defined as,
\begin{equation}
E_v=\dfrac{1}{\#\mathbb{D}_{val}}\sum_{(y_i,\bm{x_i})\in\mathbb{D}_{val}}\rho_{\tau}(y_i-\bm{x}_i^t\hat{\bm{\beta}}),
\end{equation}
where $\rho_{\tau}(\cdot)$ denotes the quantile function defined at \eqref{eq:check_function}, and $\#$ denotes the cardinal of a set. The final model error is calculated over $\mathbb{D}_{test}$ as,
\begin{equation}
E_t=\dfrac{1}{\#\mathbb{D}_{test}}\sum_{(y_i,\bm{x_i})\in\mathbb{D}_{test}}\rho_{\tau}(y_i-\bm{x}_i^t\hat{\bm{\beta}}).
\end{equation} 
Additionally, the following metrics evaluating the performance of the methods are considered:
\begin{itemize}
	\item $\norm{\hat{\bm{\beta}}-\bm{\beta}}_2$ the euclidean distance between the estimated vector and the true vector;
	\item true positive rate (TPR)$=$ P$(\hat{\beta_i}\neq0\vert\beta_i\neq 0)$;
	\item true negative rate (TNR)$=$ P$(\hat{\beta_i}=0\vert\beta_i=0)$;
	\item correct selection rate (CSR)$=$ P$(\hat{\beta}=\beta)$.
\end{itemize}

We are interested in studying the performance of the proposed models under different situations. An aspect to be analysed is the effect of an increase on the number of variables, and regarding this aspect, three cases will be considered:
	\begin{itemize}
		\item high-dimensional case with $625$ variables;
		\item high-dimensional case with $225$ variables;
		\item low dimensional case with $100$ variables.
	\end{itemize}
	
	Additionally, another important factor is the spread of the significant variables among different groups. In order to study this aspect, two cases will be considered:
	\begin{itemize}
		\item sparse distribution of significant variables: significant variables are spread among many groups, but there is no group fully formed by significant variables;
		\item dense distribution of significant variables: significant variables are concentrated into a few number of groups, fully formed by significant variables.
	\end{itemize}
	
	Varying the number and the spread of the variables, six cases will be studied:
	\\
	
	\noindent\textit{Case 1: sparse distribution of $625$ variables}\\\noindent
	There are $25$ groups of size $25$ each, a total number of $625$ variables. Among these groups, $7$ groups with $8$ significant variables each are defined, a total number of $56$ significant variables. For $l\in\{1\ldots,25\}$, coefficients inside each group are defined as,
	\begin{displaymath}
	\left\{\begin{array}{rcl}
	\beta^{l} & = & (1,2,\ldots,8,\underbrace{0,\ldots,0}_{17}),{\ }l=1,\ldots,7 \\
	\beta^{l} & = & (\underbrace{0,\ldots,0}_{25}),{\ }l=8,\ldots,25.
	\end{array} \right.
	\end{displaymath}
	\\
	
	\noindent\textit{Case 2: dense distribution of $625$ variables}\\\noindent
	There are $25$ groups of size $25$ each, a total number of $625$ variables. Among these groups, $3$ groups with $25$ significant variables each are defined, a total number of $75$ significant variables. For $l\in\{1\ldots,25\}$, coefficients inside each group are defined as,
	\begin{displaymath}
	\left\{\begin{array}{rcl}
	\beta^{l} & = & (1,2,\ldots,25),{\ }l=1,\ldots,3 \\
	\beta^{l} & = & (\underbrace{0,\ldots,0}_{25}),{\ }l=4,\ldots,25.
	\end{array} \right.
	\end{displaymath}
	\\
	
	\noindent\textit{Case 3: sparse distribution of $225$ variables}\\\noindent
	There are $15$ groups of size $15$ each, a total number of $225$ variables. Among these groups, $7$ groups with $8$ significant variables each are defined, a total number of $56$ significant variables. For $l\in\{1\ldots,15\}$, coefficients inside each group are defined as,
	\begin{displaymath}
	\left\{\begin{array}{rcl}
	\beta^{l} & = & (1,2,\ldots,8,\underbrace{0,\ldots,0}_{7}),{\ }l=1,\ldots,7 \\
	\beta^{l} & = & (\underbrace{0,\ldots,0}_{15}),{\ }l=8,\ldots,15.
	\end{array} \right.
	\end{displaymath}
	\\
	
	\noindent\textit{Case 4: dense distribution of $225$ variables}\\\noindent
	There are $15$ groups of size $15$ each, a total number of $225$ variables. Among these groups, $3$ groups with $15$ significant variables each are defined, a total number of $45$ significant variables. For $l\in\{1\ldots,15\}$, coefficients inside each group are defined as,
	\begin{displaymath}
	\left\{\begin{array}{rcl}
	\beta^{l} & = & (1,2,\ldots,15),{\ }l=1,\ldots,3 \\
	\beta^{l} & = & (\underbrace{0,\ldots,0}_{15}),{\ }l=4,\ldots,15.
	\end{array} \right.
	\end{displaymath}
	\\
	
	\noindent\textit{Case 5: sparse distribution of $100$ variables}\\\noindent
	There are $10$ groups of size $10$ each, a total number of $100$ variables. Among these groups, $5$ groups with $6$ significant variables each are defined, a total number of $30$ significant variables. For $l\in\{1\ldots,10\}$, coefficients inside each group are defined as,
	\begin{displaymath}
	\left\{\begin{array}{rcl}
	\beta^{l} & = & (1,2,\ldots,6,\underbrace{0,\ldots,0}_{4}),{\ }l=1,\ldots,5 \\
	\beta^{l} & = & (\underbrace{0,\ldots,0}_{10}),{\ }l=6,\ldots,10.
	\end{array} \right.
	\end{displaymath}
	\\
	
	\noindent\textit{Case 6: dense distribution of $100$ variables}\\\noindent
	There are $10$ groups of size $10$ each, a total number of $100$ variables. Among these groups, $3$ groups with $10$ significant variables each are defined, a total number of $30$ significant variables. For $l\in\{1\ldots,10\}$, coefficients inside each group are defined as,
	\begin{displaymath}
	\left\{\begin{array}{rcl}
	\beta^{l} & = & (1,2,\ldots,10),{\ }l=1,\ldots,3 \\
	\beta^{l} & = & (\underbrace{0,\ldots,0}_{10}),{\ }l=4,\ldots,10.
	\end{array} \right.
	\end{displaymath}
	
	We consider that \textit{Case 1} is the most representative example in further applications, and therefore it will be intensively studied here, and also in the simulations regarding the sensitivity analysis shown in the supplementary material. Each simulation example has been executed 50 times considering $100/100/5000$ observations in the train / validate / test samples, except in the low dimensional simulations (\textit{Case} $5$ and $6$) where $500/500/5000$ observations were considered. The large test sets formed by $5000$ observations help increase the stability of the results, however, models are built using train and validate sets, making the $625$ variables and $225$ variables simulations high dimensional ($p>n$). The results have been summarized in terms of the mean and standard deviation values (shown in parenthesis), and the best result from each metric is highlighted.

As it was commented in Section \ref{sec:oracle_prop}, the general tendency found in the literature regarding the weights in adaptive models is to define them based on the results of the unpenalized model,
	\begin{equation}
	\tilde{w_i}=\frac{1}{\lvert\tilde{\beta_i}\rvert^{\gamma}},
	\end{equation}
	where $w_i$  and $\tilde{\beta_i}$ correspond to the  i-\emph{th} element of vectors $\tilde{\bm{w}}$ and $\tilde{\bm{\beta}}$ respectively, $\lvert\cdot\rvert$ denotes the absolute value function, $\gamma$ is a non negative constant and $\tilde{\bm{\beta}}$ is the solution vector obtained from the unpenalized model (described, in the case of the QR framework, in equation \eqref{eq:qr}). This approach is limited just to low dimensional scenarios, where the unpenalized model can actually be solved. For this reason, in the low dimensional cases, the results of the proposed models are compared with  the results from the weights based on the unpenalized model. 


	\subsection{Simulation 1: sparse distribution of significant variables.} \label{sec:sim_sparse_t_error}
	This simulation shows the results obtained under simulation \textit{Case} $1$, considering $625$ variables, \textit{Case }$3$, considering $225$ variables and \textit{Case }$5$, considering $100$ variables. In all of them, the variables are sparsely distributed among groups, and a symmetric error from a t$(3)$ is considered. 
	
	\begin{table}[]
		\centering
		\footnotesize
		\caption{Simulation 1. Sparse distribution of variables. Considering a t(3) error.}
		\begin{tabular}{llllll}
			\toprule
			& $\norm{\hat{\bm{\beta}}-\bm{\beta}}$   & $E_t$           & CSR              & TPR              & TNR                         \\ \midrule
			\multicolumn{6}{c}{$p=625$ variables}                                                                                          \\ \midrule
			LASSO                 & $23.37$ $(4.61)$  & $7.85$ $(1.70)$ & $\bm{0.89}$ $(0.01)$  & $0.76$ $(0.07)$  & $\bm{0.90}$ $(0.01)$  \\
			SGL                   & $19.62$ $(3.28)$  & $6.29$ $(1.08)$ & $0.76$ $(0.10)$  & $0.90$ $(0.04)$  & $0.75$ $(0.12)$            \\
			AL-SGL-$pca_d$        & $17.97$ $(3.56)$  & $5.68$ $(1.13)$ & $0.83$ $(0.07)$  & $0.88$ $(0.05)$  & $0.83$ $(0.08)$            \\
			AL-SGL-$pca_1$        & $21.41$ $(2.78)$  & $6.88$ $(0.93)$ & $0.70$ $(0.10)$  & $0.90$ $(0.04)$  & $0.68$ $(0.12)$            \\
			AL-SGL-$pls_d$        & $17.60$ $(3.28)$  & $5.78$ $(1.14)$ & $0.83$ $(0.06)$  & $0.89$ $(0.04)$  & $0.83$ $(0.07)$            \\
			AL-SGL-$pls_1$        & $19.40$ $(2.99)$  & $6.23$ $(0.99)$ & $0.78$ $(0.09)$  & $0.90$ $(0.04)$  & $0.77$ $(0.10)$            \\
			ASGL-$pca_d$          & $15.19$ $(3.43)$  & $4.65$ $(1.04)$ & $0.84$ $(0.04)$  & $\bm{0.92}$ $(0.03)$  & $0.83$ $(0.04)$       \\
			ASGL-$pca_1$          & $21.38$ $(2.58)$  & $6.80$ $(0.87)$ & $0.73$ $(0.10)$  & $0.91$ $(0.04)$  & $0.71$ $(0.11)$            \\
			ASGL-$pls_d$          & $\bm{13.23}$ $(3.35)$  & $\bm{4.07}$ $(0.99)$ & $0.85$ $(0.03)$  & $0.91$ $(0.04)$  & $0.84$ $(0.04)$  \\
			ASGL-$pls_1$          & $17.56$ $(3.98)$  & $5.61$ $(1.33)$ & $0.81$ $(0.01)$  & $0.91$ $(0.04)$  & $0.80$ $(0.07)$            \\
			ASGL-$spls_d$         & $14.31$ $(3.30)$   & $4.36$ $(0.99)$ & $0.85$ $(0.03)$ & $0.92 (0.04)$ & $0.84$ $(0.04)$               \\
			ASGL-$spca_d$         & $18.05$ $(3.19)$   & $5.75$ $(1.06)$ & $0.78$ $(0.07)$ & $0.91 (0.03)$ & $0.77$ $(0.08)$               \\ \midrule
			\multicolumn{6}{c}{$p=225$ variables}                                                                                          \\ \midrule
			LASSO                 & $8.09$ $(2.48)$  & $2.66$ $(0.81)$ & $\bm{0.80}$ $(0.02)$  & $0.96$ $(0.03)$  & $\bm{0.75}$ $(0.02)$   \\
			SGL                   & $6.43$ $(2.02)$  & $2.12$ $(0.60)$ & $0.76$ $(0.06)$  & $0.98$ $(0.02)$  & $0.69$ $(0.07)$             \\
			AL-SGL-$pca_d$        & $6.66$ $(2.33)$  & $2.20$ $(0.76)$ & $0.78$ $(0.06)$  & $0.97$ $(0.03)$  & $0.71$ $(0.08)$             \\
			AL-SGL-$pca_1$        & $7.06$ $(1.98)$  & $2.30$ $(0.61)$ & $0.73$ $(0.06)$  & $0.98$ $(0.02)$  & $0.65$ $(0.09)$             \\
			AL-SGL-$pls_d$        & $6.95$ $(1.79)$  & $2.28$ $(0.56)$ & $0.77$ $(0.06)$  & $0.97$ $(0.02)$  & $0.70$ $(0.08)$             \\
			AL-SGL-$pls_1$        & $7.27$ $(2.46)$  & $2.39$ $(0.78)$ & $0.74$ $(0.06)$  & $0.98$ $(0.02)$  & $0.66$ $(0.08)$             \\
			ASGL-$pca_d$          & $5.09$ $(1.32)$  & $1.70$ $(0.38)$ & $0.73$ $(0.09)$  & $\bm{0.99}$ $(0.01)$  & $0.65$ $(0.12)$        \\
			ASGL-$pca_1$          & $7.07$ $(1.98)$  & $2.31$ $(0.62)$ & $0.75$ $(0.06)$  & $0.98$ $(0.02)$  & $0.67$ $(0.07)$             \\
			ASGL-$pls_d$          & $\bm{5.05}$ $(1.30)$  & $\bm{1.68}$ $(0.37)$ & $0.74$ $(0.09)$  & $\bm{0.99}$ $(0.02)$  & $0.66$ $(0.12)$ \\
			ASGL-$pls_1$          & $6.21$ $(1.78)$  & $2.04$ $(0.52)$ & $0.74$ $(0.05)$  & $0.98$ $(0.02)$  & $0.66$ $(0.06)$             \\ \midrule
			\multicolumn{6}{c}{$p=100$ variables}                                                                                          \\ \midrule
			LASSO                 & $0.59$ $(0.08)$  & $0.59$ $(0.01)$ & $0.79$ $(0.09)$  & $\bm{1.00}$ $(0.00)$  & $0.69$ $(0.14)$        \\
			SGL                   & $0.60$ $(0.08)$  & $0.59$ $(0.01)$ & $0.75$ $(0.11)$  & $\bm{1.00}$ $(0.00)$  & $0.64$ $(0.16)$        \\
			ASGL-$pca_d$          & $0.55$ $(0.08)$  & $0.58$ $(0.01)$ & $0.81$ $(0.10)$  & $\bm{1.00}$ $(0.00)$  & $0.73$ $(0.14)$        \\
			ASGL-$pls_d$          & $\bm{0.45}$ $(0.07)$  & $\bm{0.58}$ $(0.06)$ & $0.95$ $(0.07)$  & $\bm{1.00}$ $(0.00)$  & $0.93$ $(0.09)$ \\
			ASGL-unpenalized      & $\bm{0.45}$ $(0.07)$  & $\bm{0.58}$ $(0.05)$ & $\bm{0.96}$ $(0.07)$  & $\bm{1.00}$ $(0.00)$  & $\bm{0.95}$ $(0.07)$ \\ 
			
			\bottomrule
		\end{tabular}
		\label{tab:t_error_sparse_groups}
	\end{table}
	
	\begin{figure}[]
		\centering
		\caption{Simulation 1. Sparse distribution of $625$ variables. Considering a t(3) error. Box-plots showing the test error of the different models.}
		\includegraphics[width=12.7cm]{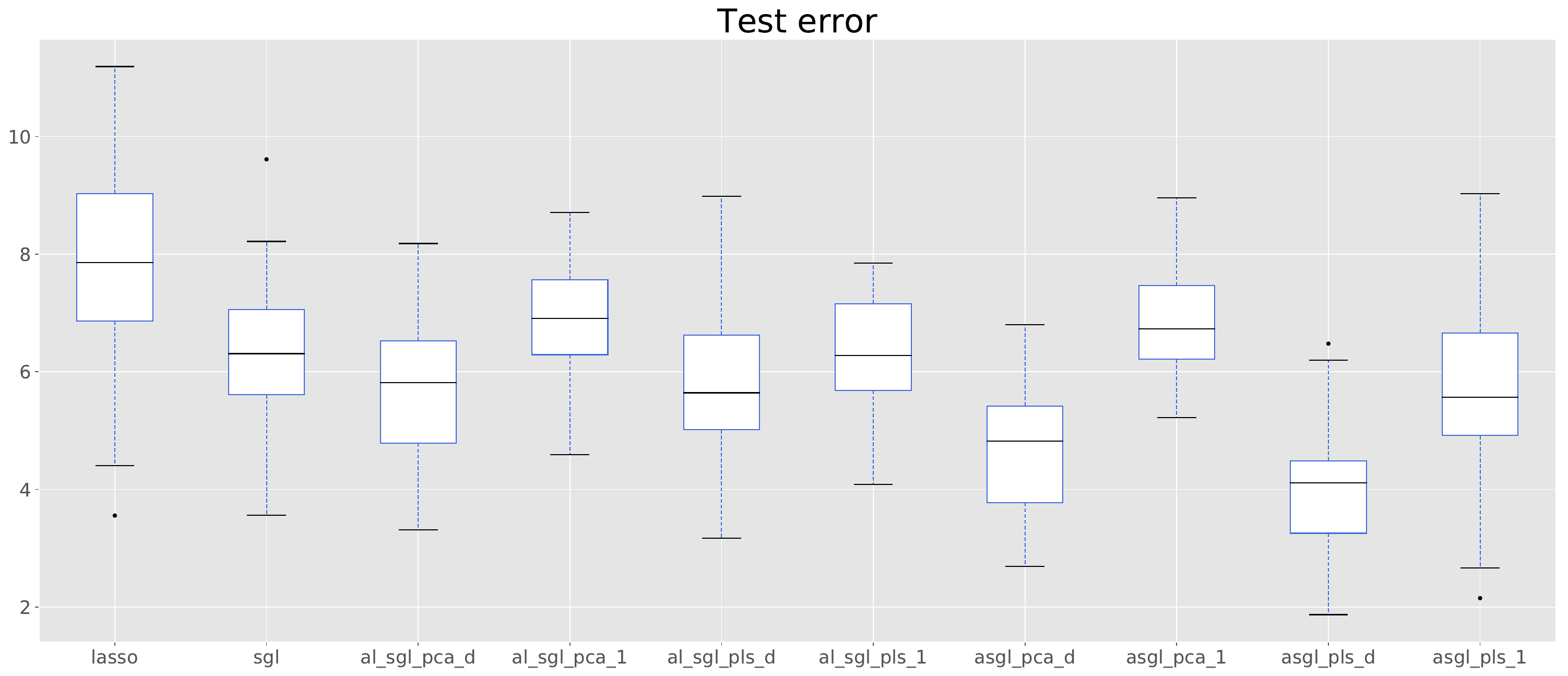}
		\label{fig:final_error_dg_t_625_sparse}
	\end{figure}
	
	\begin{figure}[]
		\centering
		\caption{Simulation 1. Sparse distribution of $225$ variables. Considering a t(3) error. Box-plots showing the test error of the different models.}
		\includegraphics[width=12.7cm]{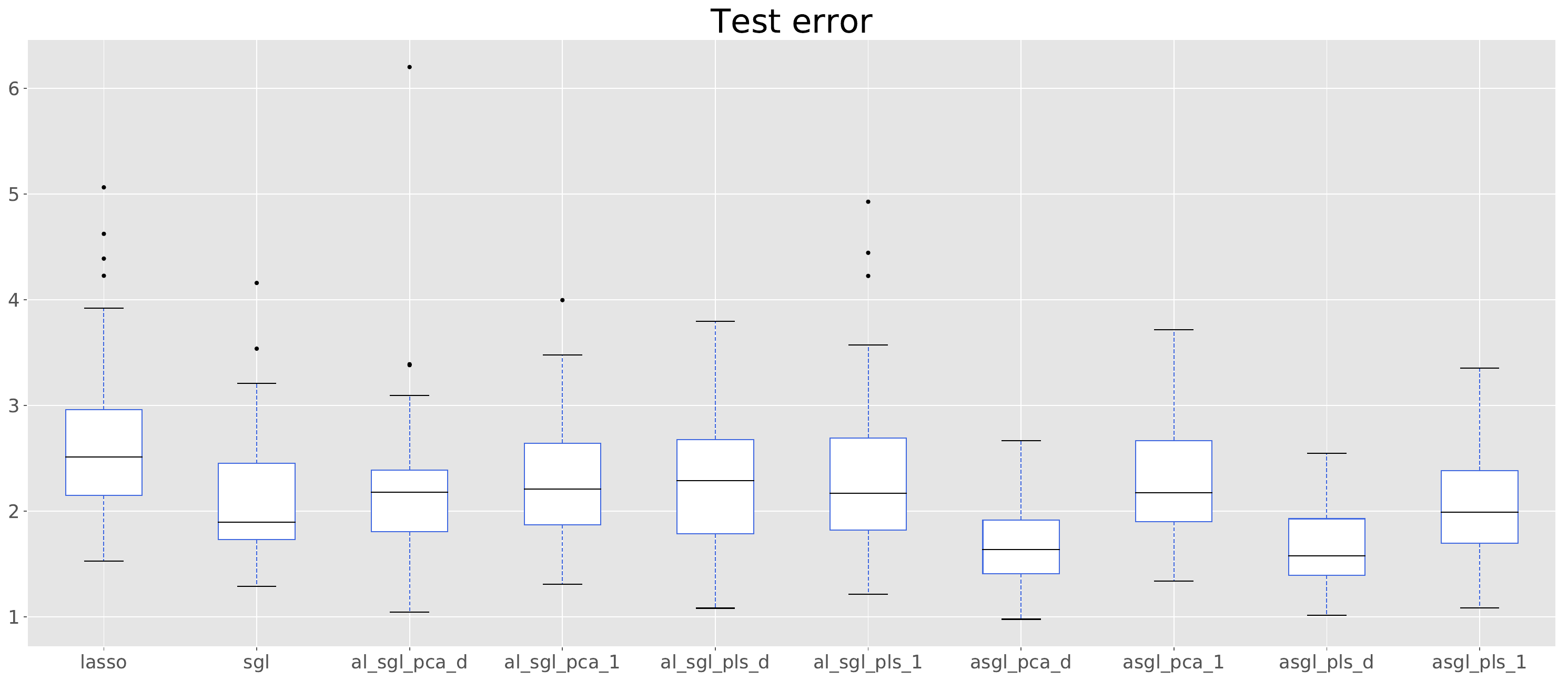}
		\label{fig:final_error_dg_t_225_sparse}
	\end{figure}
	
	Results from this simulation scheme are displayed in Table \ref{tab:t_error_sparse_groups}, which is divided into three parts related to the three \textit{Cases} under study. The first part of the table analyses \textit{Case} $1$, which considers $625$ variables. In this part, the results from LASSO and SGL are compared against the eight proposed weight calculation alternatives commented before. Additionally, the performance of sparse variations of PCA and PLS is studied. These alternatives appear denoted as $spca_d$ (from sparse PCA) and $spls_d$ (from sparse PLS). Sparse PCA was initially proposed by \citep{Zou2006} as a method that computes principal components adding a LASSO based penalization to standard PCA. This yields to principal components that are sparse linear combinations of the original variables, though are no longer orthogonal. In the same sense, \cite{Chun2010} proposed an sparse alternative to PLS. Both alternatives are studied in this simulation.The best results here are obtained by the ASGL model using $pls_d$ weights, closely followed by $spls_d$ and $pca_d$ weights. This model outperforms LASSO and SGL both in terms of the distance between predicted and true $\bm{\beta}$, and in terms of the test error $E_t$. Given that LASSO enhances individual sparsity, LASSO solutions are more sparse than the solutions obtained by the proposed models , and this is shown in the TNR values. However, LASSO offers poor results in terms of the TPR (this is, in terms of the selection of the truly significant variables). SGL shows the opposite behavior, producing solutions with large TPR values but low TNR values. Compared to these techniques, the proposed ASGL formulations achieve good variable selection results both in terms of TNR and TPR. It is worth highlighting the results achieved using the sparse PCA ($spca_d$) and sparse PLS ($spls_d$) weights alternatives. As can be seen, the performance of $spca_d$ and $spls_d$ is worse than that of $pls_d$. Our guess is that establishing a double-sparsity framework, namely, sparse components used to estimate prior weights for an adaptive sparse group LASSO, is not that beneficial, and that simple PLS may be sufficient for the weight calculation, leaving the achievement of sparse solutions to the effect of the ASGL estimator. Additionally, using sparse PCA or sparse PLS in the weight calculation requires to optimize a series of parameters related to these techniques, and then another series of parameters related to the ASGL estimator. Finding the optimal solution in such a grid of parameters can be numerically cumbersome and time-consuming.
	
	A similar behavior is observed in \textit{Case 3}, that considers $225$ variables. As before, the best results in terms of prediction accuracy are provided by ASGL $pls_d$ and $pca_d$ alternatives. Finally, the study performed in the low dimensional \textit{Case} 5 is centered on the models achieving the best results among the proposals considered, namely $pls_d$ and $pca_d$ weights, that are compared against LASSO and SGL penalizations, and against the ASGL unpenalized, which is feasible only in this low dimensional framework and that consists in estimating the weights based on a unpenalized model (as it is usually done in the literature). It is worth to remark here that the $pls_d$ alternative performs just as well as the \textit{unpenalized} one, which is a nice finding of this approach.
	
	Figures \ref{fig:final_error_dg_t_625_sparse} and \ref{fig:final_error_dg_t_225_sparse} display box-plots of the test error $E_t$ for different models in the high dimensional frameworks, showing that the spread of $E_t$ is much smaller in the ASGL $pls_d$ and $pca_d$ than in the LASSO and SGL, indicating that these models provide more stable solutions in terms of prediction accuracy.

	\subsection{Simulation 2: dense distribution of significant variables.} \label{sec:sim_dense_t_error}
	This simulation shows the results obtained under simulation \textit{Case} $2$, considering $625$ variables, \textit{Case }$4$, considering $225$ variables and \textit{Case }$6$, considering $100$ variables. In all of them, the variables are densely distributed among groups, and a symmetric error from a t$(3)$ is considered.
	
	\begin{table}[]
		\centering
		\footnotesize
		\caption{Simulation 2. Dense distribution of variables. Considering a t(3) error.}
		\begin{tabular}{llllll}
			\toprule
			& $\norm{\hat{\bm{\beta}}-\bm{\beta}}$   & $E_t$           & CSR              & TPR              & TNR             \\ \midrule
			\multicolumn{6}{c}{$p=625$ variables}                                                                              \\ \midrule
			LASSO                 & $21.00$ $(13.00)$  & $7.13$ $(4.67)$ & $\bm{0.95}$ $(0.01)$  & $0.96$ $(0.03)$  & $\bm{0.95}$ $(0.01)$ \\
			SGL                   & $6.02$ $(1.77)$  & $1.99$ $(0.56)$ & $0.82$ $(0.09)$  & $\bm{1.00}$ $(0.01)$  & $0.80$ $(0.10)$ \\
			AL-SGL-$pca_d$       & $4.32$ $(0.99)$  & $1.45$ $(0.28)$ & $0.94$ $(0.04)$  & $\bm{1.00}$ $(0.01)$  & $0.93$ $(0.05)$ \\
			AL-SGL-$pca_1$       & $7.17$ $(2.47)$  & $2.30$ $(0.75)$ & $0.72$ $(0.09)$  & $\bm{1.00}$ $(0.01)$  & $0.68$ $(0.11)$ \\
			AL-SGL-$pls_d$       & $4.81$ $(1.47)$  & $1.60$ $(0.44)$ & $0.92$ $(0.06)$  & $\bm{1.00}$ $(0.01)$  & $0.90$ $(0.07)$ \\
			AL-SGL-$pls_1$       & $5.38$ $(1.20)$  & $1.77$ $(0.57)$ & $0.87$ $(0.08)$  & $\bm{1.00}$ $(0.01)$  & $0.85$ $(0.09)$ \\
			ASGL-$pca_d$          & $\bm{3.61}$ $(0.78)$  & $\bm{1.23}$ $(0.20)$ & $0.92$ $(0.10)$  & $\bm{1.00}$ $(0.01)$  & $0.90$ $(0.12)$ \\
			ASGL-$pca_1$          & $7.60$ $(3.20)$  & $2.46$ $(1.01)$ & $0.74$ $(0.09)$  & $\bm{1.00}$ $(0.01)$  & $0.71$ $(0.11)$ \\
			ASGL-$pls_d$          & $3.85$ $(0.83)$  & $1.29$ $(0.21)$ & $0.85$ $(0.03)$  & $\bm{1.00}$ $(0.01)$  & $0.89$ $(0.13)$ \\
			ASGL-$pls_1$          & $4.17$ $(1.17)$  & $1.40$ $(0.32)$ & $0.90$ $(0.11)$  & $\bm{1.00}$ $(0.01)$  & $0.87$ $(0.09)$ \\ \midrule
			\multicolumn{6}{c}{$p=225$ variables}                                                                              \\ \midrule
			LASSO                 & $4.43$ $(1.10)$  & $1.57$ $(0.35)$ & $0.87$ $(0.03)$  & $0.99$ $(0.01)$  & $0.83$ $(0.05)$ \\
			SGL                   & $3.29$ $(0.75)$  & $1.21$ $(0.21)$ & $0.73$ $(0.13)$  & $0.99$ $(0.01)$  & $0.64$ $(0.17)$ \\
			AL-SGL-$pca_d$       & $2.88$ $(0.50)$  & $1.07$ $(0.14)$ & $0.78$ $(0.06)$  & $\bm{1.00}$ $(0.01)$  & $0.84$ $(0.11)$ \\
			AL-SGL-$pca_1$       & $3.63$ $(0.73)$  & $1.30$ $(0.22)$ & $0.61$ $(0.15)$  & $0.99$ $(0.01)$  & $0.47$ $(0.21)$ \\
			AL-SGL-$pls_d$       & $2.92$ $(0.57)$  & $1.09$ $(0.16)$ & $0.84$ $(0.12)$  & $\bm{1.00}$ $(0.01)$  & $0.78$ $(0.16)$ \\
			AL-SGL-$pls_1$       & $3.14$ $(0.65)$  & $1.16$ $(0.18)$ & $0.76$ $(0.14)$  & $\bm{1.00}$ $(0.01)$  & $0.67$ $(0.20)$ \\
			ASGL-$pca_d$          & $\bm{2.56}$ $(0.49)$  & $\bm{0.98}$ $(0.13)$ & $\bm{0.89}$ $(0.12)$  & $\bm{1.00}$ $(0.01)$  & $\bm{0.85}$ $(0.16)$ \\
			ASGL-$pca_1$          & $3.49$ $(0.79)$  & $1.25$ $(0.22)$ & $0.62$ $(0.15)$  & $\bm{1.00}$ $(0.01)$  & $0.49$ $(0.21)$ \\
			ASGL-$pls_d$          & $2.59$ $(0.43)$  & $0.99$ $(0.10)$ & $0.88$ $(0.16)$  & $\bm{1.00}$ $(0.01)$  & $0.83$ $(0.21)$ \\
			ASGL-$pls_1$          & $2.80$ $(0.53)$  & $1.05$ $(0.14)$ & $0.81$ $(0.12)$  & $\bm{1.00}$ $(0.01)$  & $0.74$ $(0.17)$ \\ \midrule
			\multicolumn{6}{c}{$p=100$ variables}                                                                             \\ \midrule
			LASSO             & $0.52$ $(0.08)$& $0.58$ $(0.01)$ & $0.82$ $(0.10)$  & $\bm{1.00}$ $(0.00)$  & $0.75$ $(0.13)$ \\
			SGL               & $0.50$ $(0.08)$  & $0.58$ $(0.01)$ & $0.74$ $(0.17)$  & $\bm{1.00}$ $(0.00)$  & $0.63$ $(0.24)$ \\
			ASGL-$pca_d$      & $0.45$ $(0.07)$  & $\bm{0.57}$ $(0.01)$ & $0.92$ $(0.11)$  & $\bm{1.00}$ $(0.00)$  & $0.88$ $(0.15)$ \\
			ASGL-$pls_d$      & $\bm{0.44}$ $(0.07)$  & $\bm{0.57}$ $(0.01)$ & $\bm{0.95}$ $(0.07)$  & $\bm{1.00}$ $(0.00)$  & $\bm{0.93}$ $(0.10)$ \\
			ASGL-unpenalized  & $0.45$ $(0.07)$  & $\bm{0.57}$ $(0.01)$ & $0.92$ $(0.12)$  & $\bm{1.00}$ $(0.00)$  & $0.89$ $(0.17)$ \\
			\bottomrule
		\end{tabular}
		\label{tab:t_error_dense_groups}
	\end{table}
	
	\begin{figure}[]
		\centering
		\caption{Simulation 2. Dense distribution of $625$ variables. Considering a t(3) error. Box-plots showing the test error of the different models.}
		\includegraphics[width=12.7cm]{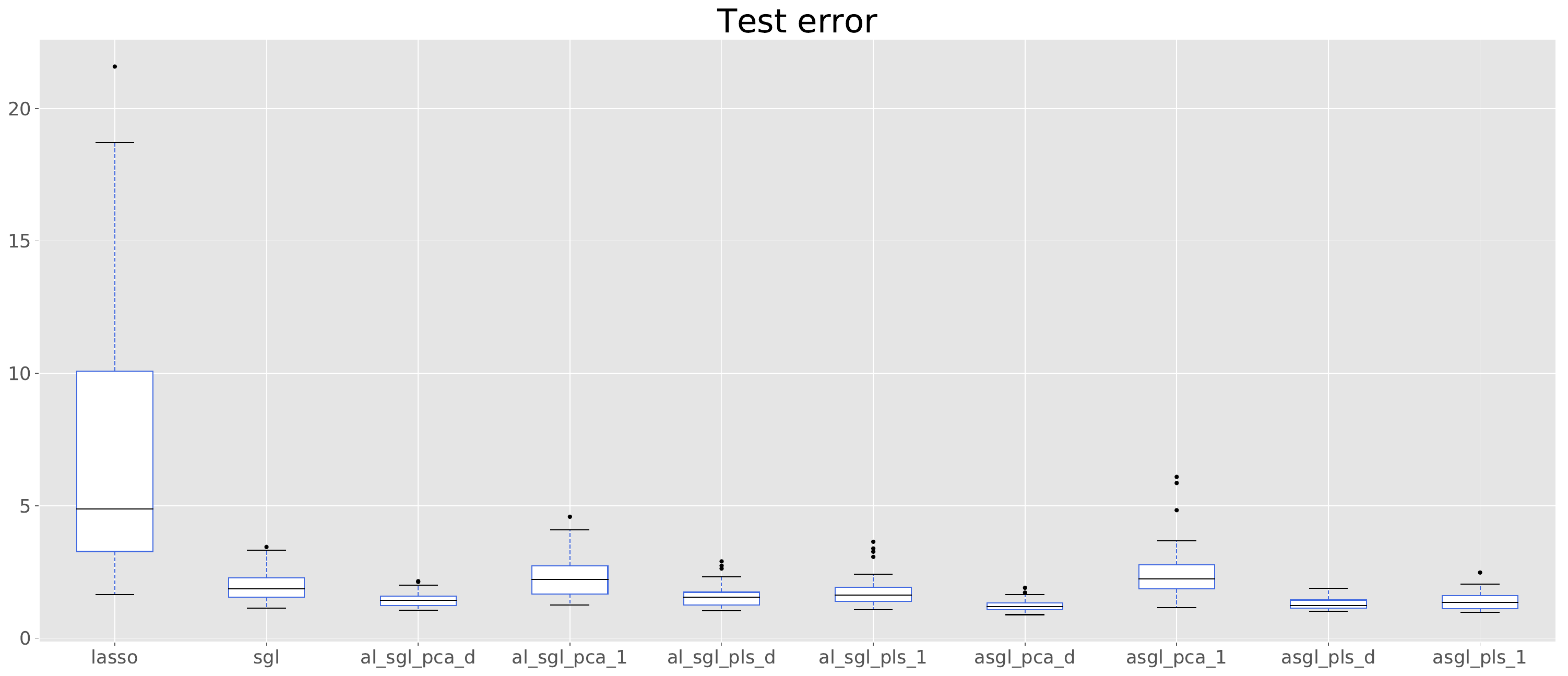}
		\label{fig:final_error_dg_t_625_dense}
	\end{figure}
	
	\begin{figure}[]
		\centering
		\caption{Simulation 2. Dense distribution of $225$ variables. Considering a t(3) error. Box-plots showing the test error of the different models.}
		\includegraphics[width=12.7cm]{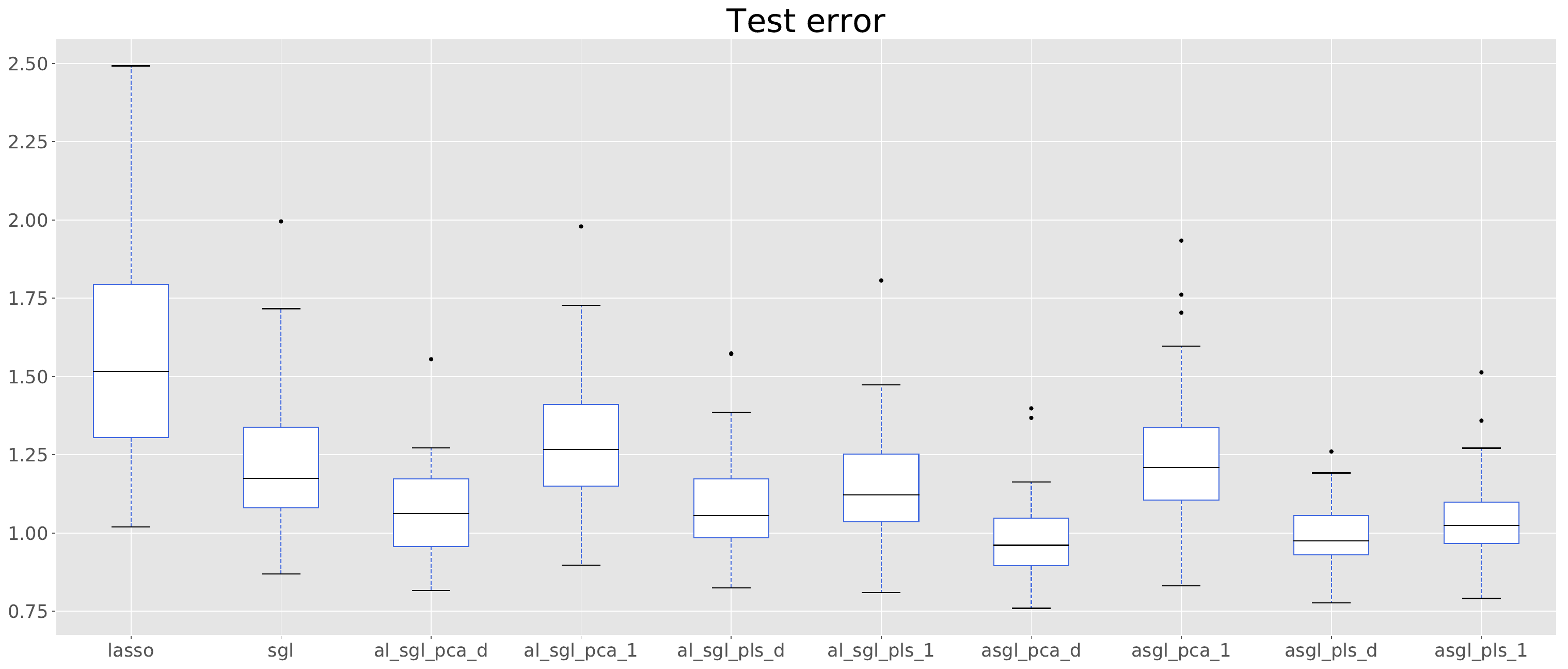}
		\label{fig:final_error_dg_t_225_dense}
	\end{figure}
	
	The results from this simulation scheme are displayed in Table \ref{tab:t_error_dense_groups}. Similar to the situation shown in the sparse distribution simulation, the ASGL model using $pls_d$ or $pca_d$ weights shows the best results in terms of the distance between predicted and true $\bm{\beta}$, and the value of $E_t$ in the high dimensional cases. These proposals offer also the best compromise between TPR and TNR. It is worth saying that under a more "compact" distribution of the significant variables in a small number of groups, the proposed methods show a great improvement in terms of prediction accuracy compared to LASSO and SGL. As before, the low dimensional \textit{case} is studied centered on the models achieving the best results among the proposals considered, $pls_d$ and $pca_d$ weights, that are compared against LASSO, SGL and ASGL \textit{unpenalized} penalizations. It can be seen here that $pls_d$ is the one achieving the best results in this framework, closely followed by $pca_d$ and \textit{unpenalized} results.
	
	Figures \ref{fig:final_error_dg_t_625_dense} and \ref{fig:final_error_dg_t_225_dense} display box-plots of test error value $E_t$ in high dimensional scenarios, showing, as in the previous simulation scheme, that ASGL models with $pls_d$ or $pca_d$ weights also provide more stable results in terms of spread. Based on previous simulations, we conclude that the best performance both in the high dimensional and low dimensional frameworks, considering sparse or dense distribution of significant variables is achieved by ASGL models with $pls_d$ or $pca_d$ weights. 

 Additionally to the simulations shown here, a comprehensive sensitivity analysis that studies the behavior of the proposed methodology under different non symmetric error distributions, when varying the powers $\gamma_1$ and 
	$\gamma_2$ entering the weights and when varying the number of PCA and PLS components chosen in the weight calculation can be found in the supplementary material.


\section{Real application}\label{sec:real_data}
The performance of the ASGL estimator is shown here using a genomic dataset first reported in \cite{Scheetz2006}. The dataset consists of $120$ twelve-week-old male offspring animals chosen for tissue harvesting from the eyes and for micro-array analysis. The dataset contains expression values from $31042$ different probe-sets (Affymetric GeneChip Rat Genome $230$ $2.0$ Array) on a logarithmic scale. As described in \cite{Huang2008b}  and \cite{Wang2012}, a two-steps preprocessing is performed, selecting, among the $31042$ probe-sets, the ones that are sufficiently expressed, and sufficiently variable. A probe is considered to be sufficiently expressed if the maximum expression value observed for that probe among the $120$ animals is greater than the $25$-\emph{th} percentile of the entire set of RMA expression values. A probe is considered to be sufficiently variable if it shows at least $2$-fold variation in the expression value among the $120$ rats. There are $18986$ probes that meet these criteria.

We study how expression level of gene TRIM32, corresponding to probe $1389163$\_at, is related to expression levels at other probes. \cite{Chiang2006} pointed out that gene TRIM32 was found to cause Bardet-Biedl syndrome, a disease of multiple organ systems including the retina.\cite[:1]{Scheetz2006} stated: “Any genetic element that can be shown to alter the expression of a specific gene or gene family known to be involved in a specific disease is itself an excellent candidate for involvement in the disease, either primarily or as a genetic modifier.” Here the sample size is $120$ (the number of animals selected for micro-array analysis), and the number of covariates (probes that pass the preprocessing steps) is $18985$. The correlation coefficients of the $18985$ probes and the probe corresponding to gene TRIM32 is calculated, and the genes in which the absolute value of the correlation exceeds $0.5$ are selected. There are $3734$ probes meeting this criteria. Finally, this dataset is standardized. Only a few genes are expected to be related to gene TRIM32, making this a high dimensional sparse problem.

From a biological perspective it is clear that genes do not work individually. The problem of grouping genes based on a medical criteria is nowadays under intense study, and it is possible to find some group structures for human genetic information based, for example, in cytogenetic positions \citep{Subramanian2005}. It is interesting to remark that groups built based on biological criteria are usually formed just by a few dozens of genes. For example, in the case of groups based on cytogenetic positions, groups  averaged $30$ genes, as stated in \cite{Simon2013}. However, these group structures are not available for all the genetic information, and to the best of our knowledge there is no genetic grouping alternative for the dataset under study here. 

We address the grouping problem from an statistical perspective, using principal components analysis to create groups of genes that are similar. It is worth to remark that in Section \ref{sec:weigths_pca} PCA was used for estimating the ASGL weights, while here it will be used for variable clustering.
\\

\noindent\textit{Variable clustering using PCA}
\begin{enumerate}
	\item Given a matrix of covariates $\bm{X}\in\mathbb{R}^{n\times p}$ as in Section \ref{sec:weigths_pca}, obtain the matrix of principal components $\bm{Q}\in\mathbb{R}^{p\times r}$ $\bm{X}\in\mathbb{R}^{n\times p}$ defined in Section \ref{sec:weigths_pca}.
	\item Consider $r$ possible groups, as many as principal components.
	\item Each principal component $\bm{q}_i\in\bm{Q}$, $i\in{1,\ldots,r}$, is a linear combination of the original variables from $\bm{X}$. Assign each original variable to the group associated to the principal component in which that variable had its maximum weight (in absolute value).
\end{enumerate}
The intuition behind this process is that variables with a large weight in the same principal component are likely to be related and should be included in the same group. 

\begin{figure}[]
	\centering
	\caption{Gene expression data of rat eye disease. Box-plot showing the sizes of the groups built using PCA.}
	\includegraphics[width=7.7cm]{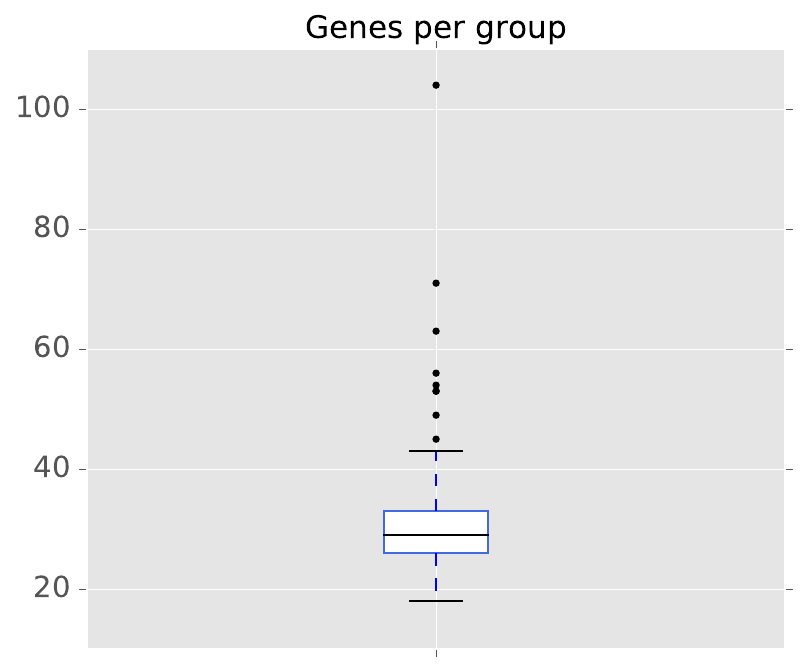}
	\label{fig:genes_per_group}
\end{figure}

In the case of the dataset used in this section, there are $120$ observations from $3734$ different genes. The maximum rank of $\bm{X}$ here is $120$, for this reason $120$ possible groups are initially considered. Each gene is assigned to the group associated to the principal component in which that gene had its maximum weight. No gene was assigned to one of the groups, and therefore $119$ groups averaging $32$ genes per group are created this way. It is worth highlighting that the average group size obtained based on this proposal is close to the expected group size in terms of the cytogenetic position. Figure \ref{fig:genes_per_group} shows a box-plot of the group sizes.

The dataset is randomly divided into $80/20/20$ train / validate / test observations and LASSO, SGL, ASGL $pls_d$ and ASGL $pca_d$ models are solved. For each model, the test error $E_t$ and the significant variables selected are obtained. This process is repeated $20$ times as a way to gain stability.

\begin{table}[]
	\centering
	\footnotesize
	\caption{Gene expression data of rat eye disease. $20$ random dataset divisions were considered. Results displayed as mean value, with standard errors in parenthesis.}
	\begin{tabular}{lll}
		\toprule
		& $E_t$          & \# Variables selected           \\ \midrule
		LASSO            & $0.34$ $(0.08)$      & $18.9$ $(15.4)  $  \\
		SGL              & $0.31$ $(0.07)$      & $189.5$ $(156.6)  $  \\
		ASGL-$pca_d$     & $\bm{0.28}$ $(0.06)$ & $56.35$ $(70.86) $  \\
		ASGL-$pls_d$     & $0.29$ $(0.06) $     & $101.7$ $(85.56)  $  \\
		\bottomrule
	\end{tabular}
	\label{tab:final_results}
\end{table}

\begin{figure}[]
	\centering
	\caption{Gene expression data of rat eye disease. $20$ random dataset divisions were considered. Box-plot showing the test error.}
	\includegraphics[width=12.7cm]{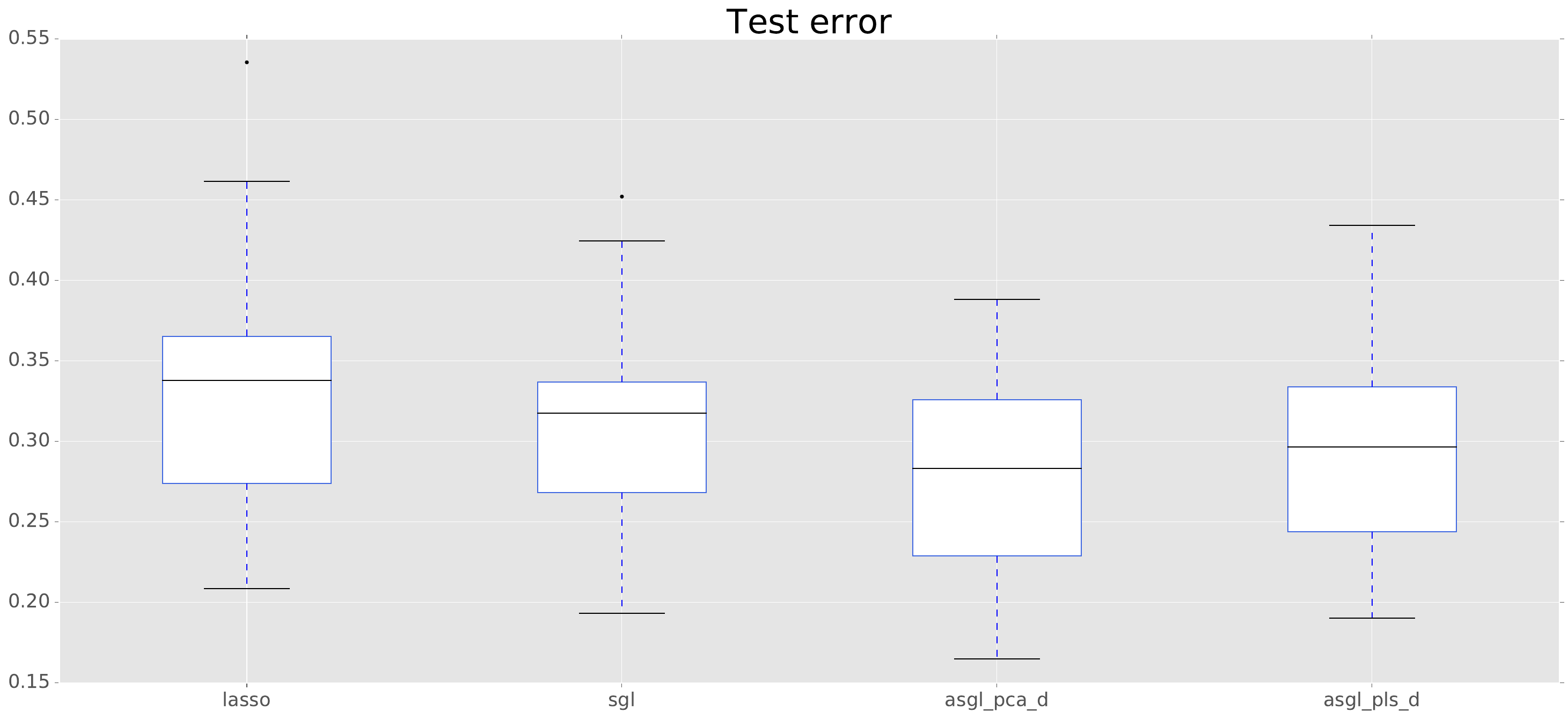}
	\label{fig:final_error_real}
\end{figure}

\begin{figure}[]
	\centering
	\caption{Gene expression data of rat eye disease. $20$ random dataset divisions were considered. Box-plot showing the number of significant genes.}
	\includegraphics[width=12.7cm]{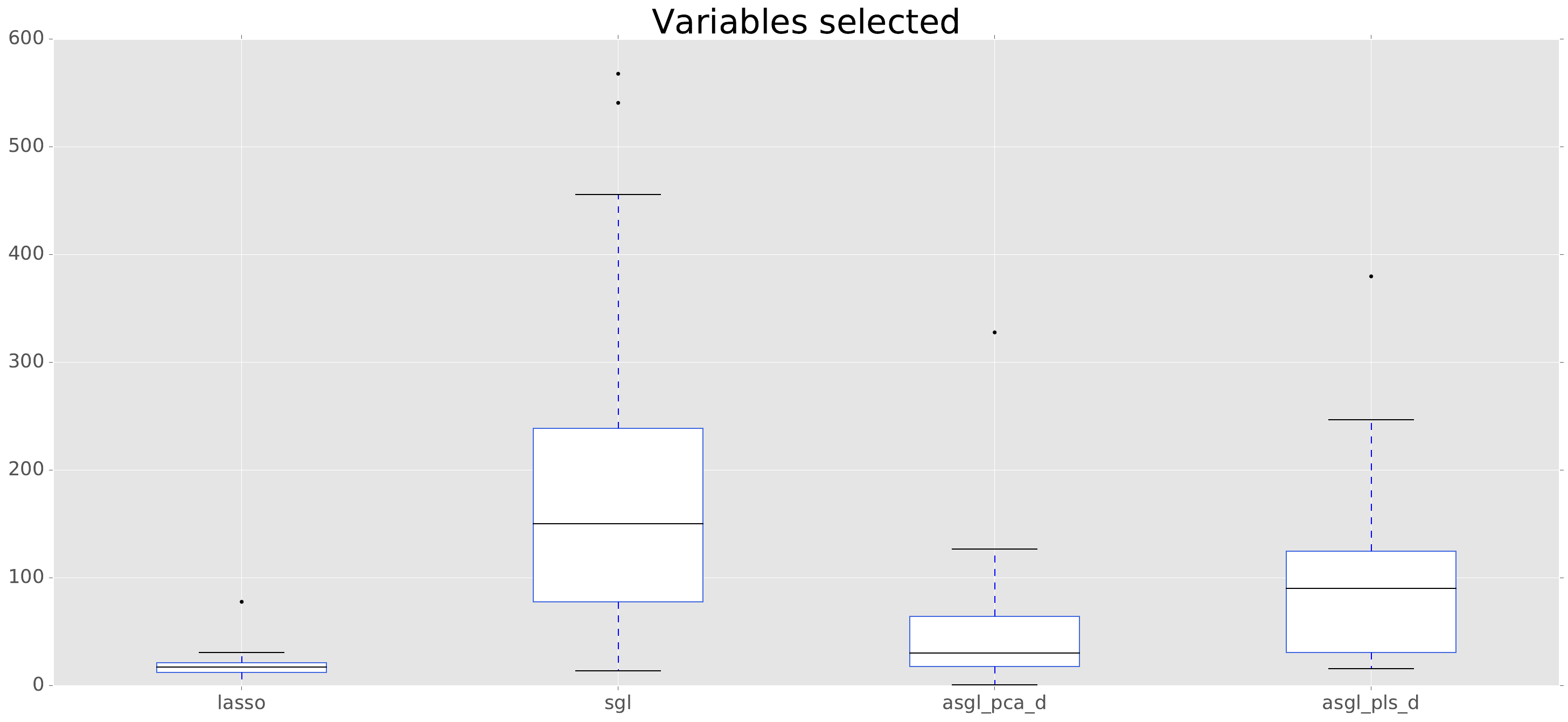}
	\label{fig:non_zero_pred_beta_real}
\end{figure}

The results obtained are shown in Table \ref{tab:final_results}. The best results in terms of the test error are obtained by the proposed ASGL models. LASSO offers a test error approximately $20\%$ greater while SGL test error is $11\%$ greater. Figure \ref{fig:final_error_real} displays box-plots of the test error $E_t$, showing that the spread of $E_t$ is also smaller in the proposed ASGL models providing more stable results. Figure \ref{fig:non_zero_pred_beta_real} displays box-plots of the number of genes each model selected as significant. The LASSO is the one offering more sparse solutions, using only $19$ variables (in mean) per model. SGL is the one using the largest number of variables, approximately $190$, and also the one with the largest variability in this metric. Both ASGL $pca_d$ and ASGL $pls_d$ selected a smaller number of variables than SGL but still larger than LASSO, and they achieve the best prediction results of the four models. 

\begin{figure}[]
	\centering
	\caption{Gene expression data of rat eye disease. $20$ random dataset divisions were considered. Heatmap showing the probability of being a significant gene. Each row represents a model and each column represents a gene.}
	\includegraphics[width=12cm]{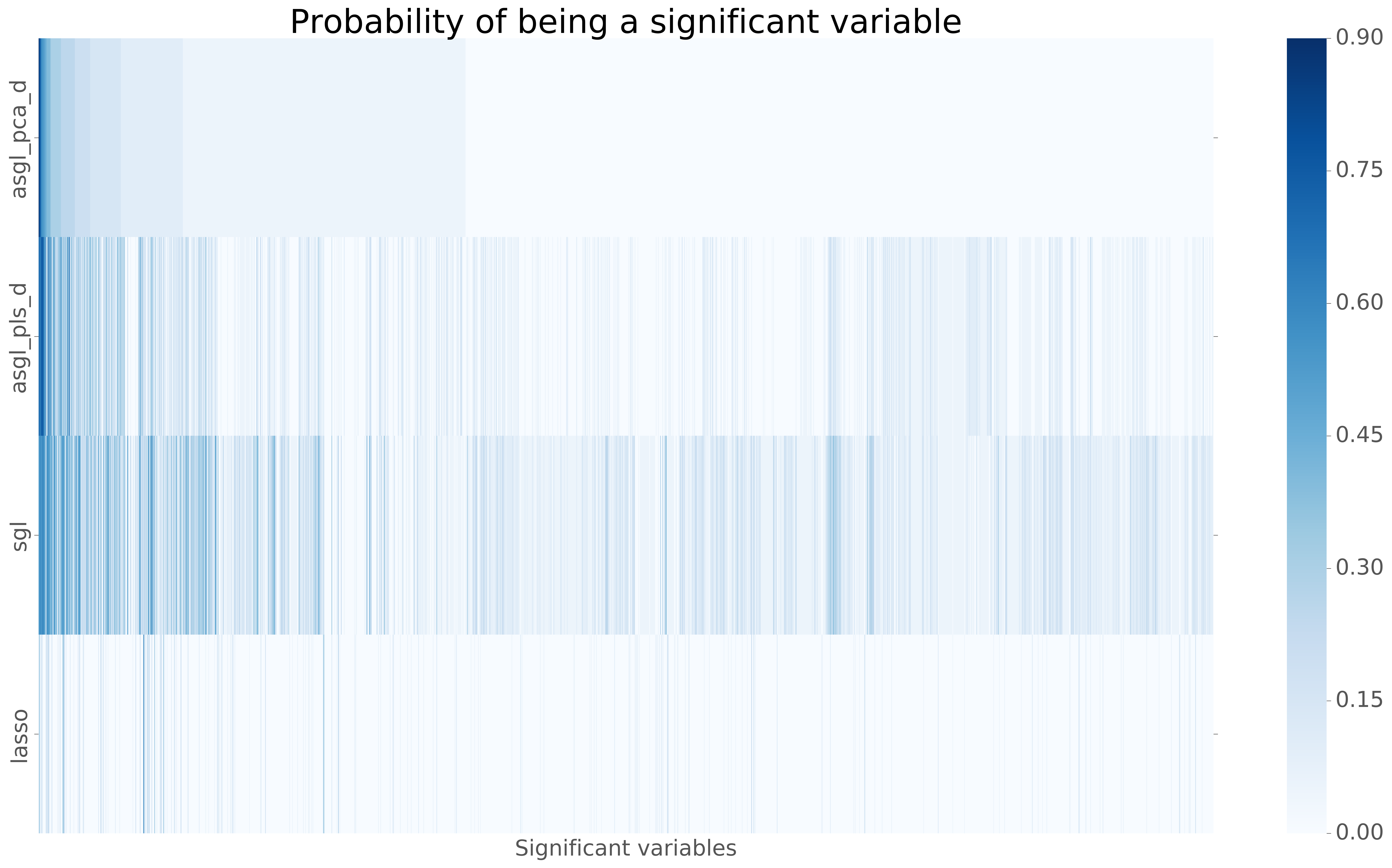}
	\label{fig:heatmap}
\end{figure}

Given that we have the results obtained from $20$ repetitions, it is possible to count the number of times each gene has been selected as significant by one of the models in any of the repetitions. Dividing this number by the total number of repetitions, a sort of "probability of being a significant gene" associated to each gene for each model considered is obtained. Out of the $3734$ genes in the dataset, $1612$ genes were selected at least one time by any of the models in any of the repetitions (the majority being selected by SGL models). Figure \ref{fig:heatmap} shows the probability of being a significant gene for these $1612$ variables and for each model. Rows represent the different models considered and columns represent each gene. Genes are sorted based on the probabilities obtained in the ASGL model with $pca_d$ weights. 

Considering a probability threshold of $0.5$, only $1$ gene in the LASSO models reach a probability of significance above the threshold, showing no stability on the gene selection along the $20$ repetitions, and anticipating problems with possible further biological interpretation of the statistical results. In the case of the SGL model, $35$ genes are above the probability threshold, being $0.6$ the maximum probability achieved. On the other hand, the ASGL model with $pls_d$ weights includes $17$ genes with probabilities above the threshold with a maximum probability value of $0.75$, and the ASGL model with $pca_d$ weights has $9$ genes above the probability threshold with a maximum probability value of $0.9$, showing more stability on the selection along the $20$ repetitions and possibly better biological interpretation of the results than the other models.

Results displayed in Table \ref{tab:final_results} and Figure \ref{fig:heatmap} have been obtained using estimators of the median of the response variable, however, it can be interesting to compare the genes selected at different quantiles. For this reason, the process described above is repeated and LASSO, SGL, ASGL $pls_d$ and ASGL $pca_d$ models are solved for quantile levels $\tau=0.3$ and $\tau=0.7$, obtaining probabilities of being a significant gene for each quantile level and each model. Considering a probability threshold of $0.5$, Table \ref{tab:common_genes} show the number of genes above the probability threshold for each quantile, and also the number of genes in the same model that have been selected along the different quantile levels.

\begin{table}[]
	\centering
	\footnotesize
	\caption{Gene expression data of rat eye disease. $20$ random dataset divisions were considered. Number of genes above the probability threshold for different quantile levels.}
	\begin{tabular}{lllll}
		\toprule
		\multicolumn{5}{c}{Number of genes above the probability threshold}       \\ \midrule
		& $\tau=0.3$ & $\tau=0.5$ & $\tau=0.7$ & Three quantiles \\ \midrule
		LASSO            & $0 $       & $1 $       & $1 $       & $0$             \\
		SGL              & $19$       & $35$       & $17$       & $0$             \\
		ASGL-$pca_d$     & $23$       & $9 $       & $17$       & $7$             \\
		ASGL-$pls_d$     & $41$       & $17$       & $37$       & $9$             \\
		\bottomrule
	\end{tabular}
	\label{tab:common_genes}
\end{table}

The LASSO model shows no stability on the variable selection, having only one gene above the threshold for $\tau=0.5$ and $\tau=0.7$, and no gene with probability of being significant above $0.5$ on the three quantiles simultaneously. The SGL shows some stability across the $20$ repetitions considering each quantile independently, but when considering all the quantiles simultaneously it has no gene above the probability threshold.  On the other hand, in the case of the ASGL $pls_d$ model, $9$ genes had a probability of being significant greater than $0.5$ in the $3$ quantiles, and in the case of the ASGL $pca_d$ models, $7$ genes fulfilled this, showing more robust results than the other estimators.

We conclude that the best results in this real dataset study are provided by the ASGL model with $pca_d$ weights, given that this model is the one with the smallest prediction error and showing great stability on the gene selection.


\section{Computational aspects}\label{sec:comp}
All the simulations and data analysis commented in Sections \ref{sec:sym_simetric_error}, and \ref{sec:real_data} and in the supplementary material were run in a cluster node with two 
Intel (R) Xeon(R) CPU E5-2630 v3 (2.4GHz, 20MB Smart Cache) processors, with 32Gb of RAM memory running CentOS 6.5 Final (Rocks 6.1.1 Sand Boa). The computation itself has been developed in Python 2.7.15 (Anaconda Inc.). All the optimization problems have been solved using the CVXPY optimization framework for Python \citep{Diamond2016} and the open source solver ECOS \citep{Domahidi2013}.


\section{Conclusion}\label{sec:conclusion}
In this paper the definition of the SGL estimator has been extended to the QR framework. A new estimator for quantile regression based on the usage of adaptive weights, the adaptive sparse group LASSO in quantile regression has also been proposed. As shown in Section \ref{sec:oracle_prop}, adaptive penalizations are typically centered on the  study of the oracle property in both asymptotic and double asymptotic frameworks. A key step on the demonstration of this property is the usage of an initial $\sqrt{n}$-consistent estimator that is usually the result of a nonpenalized model. However, this definition limits the usage of adaptive estimators to low dimensional scenarios. As a solution to this problem, four weight calculation alternatives that can be used in high dimensional scenarios when working with adaptive estimators have been proposed. Section \ref{sec:oracle_proerties_wc} conjectures about the relation between these alternatives and the oracle property. Additionally, the performance of the proposed alternatives have been analyzed in a set of synthetic data scenarios that includes high dimensional and low dimensional examples and symmetric error distributions (Section \ref{sec:sym_simetric_error}). Moreover, a thorough sensitivity analysis studying the behavior of the estimator under different error distributions, and under changes in parameter values has been performed in the supplementary material. The performance of the proposed work is also studied in a real high dimensional dataset including gene expression values of rat eye disease. Previous synthetic data analysis showed that the ASGL estimator is a competitive option in both high and low dimensional scenarios, especially when the adaptive weights are calculated based on subsets of PCA or PLS components. However, when dealing with the real dataset, the ASGL $pca_d$ estimator achieved better results in terms of prediction error and stability of the variables selected. For this reason we conclude that the ASGL $pca_d$ provides the best results among the options proposed in this work. 

This work has risen some questions that will require further investigation. One interesting problem is the optimization of the hyper-parameters. In this work we make use of grid-search, but it is worth commenting that new hyper-parameter tuning alternatives have appeared in recent years \citep{Laria2019}, and it can be interesting to investigate the usage of this or other options in the optimization of the parameters of the models introduced in this work. 

Section \ref{sec:oracle_proerties_wc} has shown some concluding remarks related to the oracle property of the $pca_d$ weight calculation alternative. The $pls_d$ alternative based on PLS, however, is more complex and will require further research. In any case, it is worth mentioning the interesting work performed by \cite{Chun2010}, that studies the consistency of the PLS estimator in the asymptotic and double asymptotic frameworks, reaching the conclusion (in \textit{Theorem 1}) that given some previous assumptions, if $\frac{p}{n}\rightarrow0$, then
$$\left\|\beta^{PLS}-\beta\right\|_2\rightarrow0\textup{{\ } in probability}.$$
This result would prove the consistency of the estimator, but It would not be enough for proving the $\sqrt{n}$-consistency, for this reason, we consider that the asymptotic property of the $pls_d$ alternative is a topic for future work.

Finally, simulations from Section \ref{sec:sym_simetric_error} have studied different model formulations, including (suggested by a referee) the usage of sparse PCA and sparse PLS in the weight calculation process. The simulations showed that this alternative did not yield to better results than the non sparse PCA or PLS alternatives, but it can be interesting to study other sparse techniques.


\section{Acknowledgments}
We appreciate the work of the referees that has contributed to substantially improve the scientific contributions of this work.
In this research we have made use of Uranus, a supercomputer cluster located at University Carlos III of Madrid and funded jointly by EU-FEDER funds and by the Spanish Government via the National Projects No. UNC313-4E-2361, No. ENE2009-12213- C03-03, No. ENE2012-33219 and No. ENE2015-68265-P. This research was partially supported by research grants and Project ECO2015-66593-P from Ministerio de Economía, Industria y Competitividad, Project MTM2017-88708-P from Ministerio de Economía y Competitividad, FEDER funds and Project  IJCI-2017-34038 from Agencia Estatal de Investigación, Ministerio de Ciencia, Innovación y Universidades.

\bibliography{references}

@article{Tibshirani1996,
author = {Tibshirani, Robert},
doi = {10.2307/2346178},
journal = {Journal of the Royal Statistical Society. Series B (Methodological)},
number = {1},
pages = {267--288},
publisher = {WileyRoyal Statistical Society},
title = {{Regression Shrinkage and Selection via the Lasso}},
volume = {58},
year = {1996}
}

@article{Ghosh2011,
author = {Ghosh, Samiran},
doi = {10.1007/s11222-010-9181-4},
journal = {Statistics and computing},
pages = {451--462},
title = {{On the grouped selection and model complexity of the adaptive elastic net}},
url = {https://link.springer.com/content/pdf/10.1007{\%}2Fs11222-010-9181-4.pdf},
volume = {21},
year = {2011}
}

@article{Yuan2006,
author = {Yuan, Ming and Lin, Yi},
journal = {Journal of the Royal Statistical Society. Series B (Methodological)},
number = {1},
pages = {49--67},
title = {{Model selection and estimation in regression with grouped variables}},
volume = {68},
year = {2006}
}

@article{Fan2004,
author = {Fan, Jianqing and Peng, Heng},
doi = {10.1214/009053604000000256},
issn = {00905364},
journal = {Annals of Statistics},
number = {3},
pages = {928--961},
publisher = {Institute of Mathematical Statistics},
title = {{Nonconcave penalized likelihood with a diverging number of parameters}},
volume = {32},
year = {2004}
}

@article{Friedman2010,
archivePrefix = {arXiv},
arxivId = {1001.0736},
author = {Friedman, J. and Hastie, T. and Tibshirani, R.},
doi = {10.1111/biom.12292},
eprint = {1001.0736},
isbn = {0006-341X},
issn = {15410420},
journal = {ArXiv:1001.0736},
pages = {1--8},
pmid = {25732839},
title = {{A note on the group lasso and a sparse group lasso}},
url = {http://arxiv.org/abs/1001.0736},
year = {2010}
}

@inproceedings{Domahidi2013,
author = {Domahidi, Alexander and Chu, Eric and Boyd, Stephen},
booktitle = {European Control Conference (ECC)},
doi = {10.0/Linux-x86_64},
isbn = {9783952417348},
title = {{ECOS: An SOCP Solver for Embedded Systems}},
year = {2013}
}

@article{Fan2001,
author = {Fan, Jianqing and Li, Runze},
doi = {10.2307/3085904},
isbn = {0162-1459},
issn = {0162-1459},
journal = {Journal of the American Statistical Association},
number = {456},
pages = {1348--1360},
pmid = {21796725},
publisher = {Taylor {\&} Francis, Ltd.American Statistical Association},
title = {{Variable Selection via Nonconcave Penalized Likelihood and Its Oracle Properties}},
volume = {96},
year = {2001}
}

@article{Chiang2006,
author = {Chiang, A. P. and Beck, J. S. and Yen, H.-J. and Tayeh, M. K. and Scheetz, T. E. and Swiderski, R. E. and Nishimura, D. Y. and Braun, T. A. and Kim, K.-Y. A. and Huang, J. and Elbedour, K. and Carmi, R. and Slusarski, D. C. and Casavant, T. L. and Stone, E. M. and Sheffield, V. C.},
doi = {10.1073/pnas.0600158103},
journal = {Proceedings of the National Academy of Sciences},
month = {apr},
number = {16},
pages = {6287--6292},
pmid = {16606853},
title = {{Homozygosity mapping with SNP arrays identifies TRIM32, an E3 ubiquitin ligase, as a Bardet-Biedl syndrome gene (BBS11)}},
volume = {103},
year = {2006}
}

@article{Loh2017,
archivePrefix = {arXiv},
arxivId = {1501.00312},
author = {Loh, Po Ling},
doi = {10.1214/16-AOS1471},
eprint = {1501.00312},
issn = {00905364},
journal = {Annals of Statistics},
number = {2},
pages = {866--896},
title = {{Statistical consistency and asymptotic normality for high-dimensional robust m-estimators}},
volume = {45},
year = {2017}
}

@article{Li2008,
author = {Li, Youjuan and Zhu, Ji},
doi = {10.1198/106186008X289155},
journal = {Journal of Computational and Graphical Statistics},
month = {mar},
number = {1},
pages = {1--23},
publisher = {Taylor {\&} Francis},
title = {{L1- -Norm Quantile Regression}},
volume = {17},
year = {2008}
}

@article{Huang2008b,
author = {Huang, Jian and Ma, Shuangge and Zhang, Cun-Hui},
journal = {Statistica Sinica},
number = {374},
pages = {1--28},
title = {{Adaptive Lasso for Sparse High-dimensional Regression}},
volume = {1},
year = {2008}
}

@article{Zhou2010,
archivePrefix = {arXiv},
arxivId = {1006.2871},
author = {Zhou, Nengfeng and Zhu, Ji},
eprint = {1006.2871},
journal = {Statistics and Its Interface},
pages = {557--574},
title = {{Group Variable Selection via a Hierarchical Lasso and Its Oracle Property}},
url = {http://arxiv.org/abs/1006.2871},
volume = {3},
year = {2010}
}

@article{YahyaAlgamal2019,
author = {{Yahya Algamal}, Zakariya and {Hisyam Lee}, Muhammad},
doi = {10.1007/s11634-018-0334-1},
journal = {Advances in Data Analysis and Classification},
pages = {753--771},
title = {{A two-stage sparse logistic regression for optimal gene selection in high-dimensional microarray data classification}},
url = {https://doi.org/10.1007/s11634-018-0334-1},
volume = {13},
year = {2019}
}

@article{Scheetz2006,
author = {Scheetz, T. E. and Kim, K.-Y. A. and Swiderski, R. E. and Philp, A. R. and Braun, T. A. and Knudtson, K. L. and Dorrance, A. M. and DiBona, G. F. and Huang, J. and Casavant, T. L. and Sheffield, V. C. and Stone, E. M.},
doi = {10.1073/pnas.0602562103},
journal = {Proceedings of the National Academy of Sciences},
month = {sep},
number = {39},
pages = {14429--14434},
pmid = {16983098},
title = {{Regulation of gene expression in the mammalian eye and its relevance to eye disease}},
volume = {103},
year = {2006}
}

@article{Kim2008,
author = {Kim, Yongdai and Choi, Hosik and Oh, Hee Seok},
doi = {10.1198/016214508000001066},
issn = {01621459},
journal = {Journal of the American Statistical Association},
number = {484},
pages = {1665--1673},
title = {{Smoothly clipped absolute deviation on high dimensions}},
volume = {103},
year = {2008}
}

@article{Subramanian2005,
author = {Subramanian, A. and Tamayo, P. and Mootha, V. K. and Mukherjee, S. and Ebert, B. L. and Gillette, M. A. and Paulovich, A. and Pomeroy, S. L. and Golub, T. R. and Lander, E. S. and Mesirov, J. P.},
doi = {10.1073/pnas.0506580102},
journal = {Proceedings of the National Academy of Sciences},
month = {oct},
number = {43},
pages = {15545--15550},
pmid = {16199517},
title = {{Gene set enrichment analysis: A knowledge-based approach for interpreting genome-wide expression profiles}},
volume = {102},
year = {2005}
}

@book{Huber2009,
address = {Hoboken, NJ, USA},
author = {Huber, Peter J. and Ronchetti, Elvezio M.},
booktitle = {Robust Statistics: Second Edition},
doi = {10.1002/9780470434697},
isbn = {9780470434697},
month = {feb},
pages = {1--354},
publisher = {wiley},
series = {Wiley Series in Probability and Statistics},
title = {{Robust Statistics: Second Edition}},
url = {http://doi.wiley.com/10.1002/9780470434697},
year = {2009}
}

@article{Poignard2018,
author = {Poignard, Benjamin},
doi = {10.1007/s10463-018-0692-7},
issn = {15729052},
journal = {Annals of the Institute of Statistical Mathematics},
title = {{Asymptotic theory of the adaptive Sparse Group Lasso}},
year = {2018}
}

@article{Diamond2016,
archivePrefix = {arXiv},
arxivId = {1603.00943},
author = {Diamond, Steven and Boyd, Stephen},
eprint = {1603.00943},
journal = {arXiv:1603.00943},
month = {mar},
title = {{CVXPY: A Python-Embedded Modeling Language for Convex Optimization}},
year = {2016}
}

@article{Wang2012,
author = {Wang, Lan and Wu, Yichao and Li, Runze},
doi = {10.1080/01621459.2012.656014},
isbn = {6176321972},
issn = {01621459},
journal = {Journal of the American Statistical Association},
number = {497},
pages = {214--222},
pmid = {1000000221},
title = {{Quantile regression for analyzing heterogeneity in ultra-high dimension}},
volume = {107},
year = {2012}
}

@article{Ciuperca2019,
author = {Ciuperca, Gabriela},
doi = {10.1007/s00362-016-0832-1},
issn = {09325026},
journal = {Statistical Papers},
month = {feb},
number = {1},
pages = {173--197},
publisher = {Springer Berlin Heidelberg},
title = {{Adaptive group LASSO selection in quantile models}},
volume = {60},
year = {2019}
}

@article{Wu2009,
author = {Wu, Yichao and Liu, Yufeng},
journal = {Statistica Sinica},
number = {2},
pages = {801--817},
title = {{Variable selection in quantile regression}},
volume = {19},
year = {2009}
}

@article{Wright2010,
author = {Wright, John and Ma, Yi and Mairal, Julien and Sapiro, Guillermo and Huang, Thomas S. and Yan, Shuicheng},
doi = {10.1109/JPROC.2010.2044470},
issn = {0018-9219},
journal = {Proceedings of the IEEE},
month = {jun},
number = {6},
pages = {1031--1044},
title = {{Sparse Representation for Computer Vision and Pattern Recognition}},
volume = {98},
year = {2010}
}

@article{Zhao2014,
author = {Zhao, Weihua and Zhang, Riquan and Liu, Jicai},
doi = {10.1080/02664763.2014.888541},
issn = {0266-4763},
journal = {Journal of Applied Statistics},
month = {aug},
number = {8},
pages = {1658--1677},
title = {{Sparse group variable selection based on quantile hierarchical Lasso}},
volume = {41},
year = {2014}
}

@article{Ciuperca2017,
author = {Ciuperca, Gabriela},
doi = {10.1080/15598608.2016.1258601},
issn = {15598616},
journal = {Journal of Statistical Theory and Practice},
month = {jan},
number = {1},
pages = {107--125},
publisher = {Taylor {\&} Francis},
title = {{Adaptive fused LASSO in grouped quantile regression}},
volume = {11},
year = {2017}
}

@article{Zou2006,
author = {Zou, Hui and Hastie, Trevor and Tibshirani, Robert},
doi = {10.1198/106186006X113430},
journal = {Journal of Computational and Graphical Statistics},
number = {2},
pages = {265--286},
title = {{Sparse Principal Component Analysis}},
volume = {15},
year = {2006}
}

@inproceedings{Chatterjee2011,
author = {Chatterjee, Soumyadeep and {Banerjee, Arindam}, Snigdhanshu and Ganguly, Auroop R.},
booktitle = {2011 IEEE 11th International Conference on Data Mining Workshops},
doi = {10.1109/ICDMW.2011.155},
isbn = {978-1-4673-0005-6},
month = {dec},
pages = {1--8},
publisher = {IEEE},
title = {{Sparse Group Lasso for Regression on Land Climate Variables}},
year = {2011}
}

@article{Koenker1978,
author = {Koenker, Roger and Bassett, Gilbert},
doi = {10.2307/1913643},
issn = {00129682},
journal = {Econometrica},
month = {jan},
number = {1},
pages = {33--50},
publisher = {The Econometric Society},
title = {{Regression Quantiles}},
volume = {46},
year = {1978}
}

@article{Simon2013,
author = {Simon, Noah and Friedman, Jerome and Hastie, Trevor and Tibshirani, Robert},
doi = {10.1080/10618600.2012.681250},
issn = {10618600},
journal = {Journal of Computational and Graphical Statistics},
month = {apr},
number = {2},
pages = {231--245},
publisher = {Taylor {\&} Francis Group},
title = {{A sparse-group lasso}},
volume = {22},
year = {2013}
}

@article{Nardi2008,
author = {Nardi, Yuval and Rinaldo, Alessandro},
doi = {10.1214/08-EJS200},
issn = {19357524},
journal = {Electronic Journal of Statistics},
number = {0},
pages = {605--633},
publisher = {The Institute of Mathematical Statistics and the Bernoulli Society},
title = {{On the asymptotic properties of the group lasso estimator for linear models}},
volume = {2},
year = {2008}
}

@book{Koenker2005,
author = {Koenker, Roger},
isbn = {0521338255},
pages = {366},
publisher = {Cambridge university Press},
title = {{Quantile Regression}},
year = {2005}
}

@article{Huang2008,
author = {Huang, Jian and Horowitz, Joel L. and Ma, Shuangge},
doi = {10.1214/009053607000000875},
issn = {00905364},
journal = {The Annals of Statistics},
month = {apr},
number = {2},
pages = {587--613},
title = {{Asymptotic properties of bridge estimators in sparse high-dimensional regression models}},
volume = {36},
year = {2008}
}

@article{Chun2010,
author = {Chun, Hyonho and Keleş, S{\"{u}}nd{\"{u}}z},
doi = {10.1111/j.1467-9868.2009.00723.x},
issn = {13697412},
journal = {Journal of the Royal Statistical Society. Series B: Statistical Methodology},
month = {jan},
number = {1},
pages = {3--25},
pmid = {20107611},
publisher = {Wiley-Blackwell},
title = {{Sparse partial least squares regression for simultaneous dimension reduction and variable selection}},
volume = {72},
year = {2010}
}

@article{Laria2019,
author = {Laria, Juan C. and Aguilera-Morillo, M. Carmen and Lillo, Rosa E.},
doi = {10.1080/10618600.2019.1573687},
journal = {Journal of Computational and Graphical Statistics},
month = {feb},
pages = {1--21},
title = {{An iterative sparse-group lasso}},
year = {2019}
}

@article{Zou2006a,
author = {Zou, Hui},
doi = {10.1198/016214506000000735},
issn = {0162-1459},
journal = {Journal of the American Statistical Association},
month = {dec},
number = {476},
pages = {1418--1429},
publisher = {Taylor {\&} Francis},
title = {{The Adaptive Lasso and Its Oracle Properties}},
volume = {101},
year = {2006}
}
\end{document}